\documentstyle[12pt,epsf]{article}

\def\doublespace
{\smallskipamount=7.5pt plus2pt minus2pt \medskipamount=15pt plus4pt
minus4pt \bigskipamount=30pt plus8pt minus8pt \normalbaselineskip=30pt
plus0pt minus0pt
\normallineskip=2pt \normallineskiplimit=0pt \jot=7.5pt
{\def\smallskip {\vskip\smallskipamount}} {\def\medskip
{\vskip\medskipamount}} {\def\bigskip {\vskip\bigskipamount}}
{\setbox\strutbox=\hbox{\vrule height21.0pt depth9.0pt width 0pt}}
\parskip 15.0pt \normalbaselines}

\def\mins{\mbox{$\overline{MS}$}}
\def\mz{\mbox{$m_Z$}}

\def\m12{\mbox{$M_{1/2}$}}
\def\ms0{\mbox{$M_0$}}
\def\msw{\mbox{$M_2$}}
\def\msg{\mbox{$M_3$}}
\def\msh{\mbox{$\tilde{m}_h$}}
\def\mww{\mbox{$M_{1/2}^2$}}
\def\mss0{\mbox{$M_0^2$}}

\def\msq{\mbox{$\tilde{m}_{q_L}$}}

\def\msqe{\mbox{$\tilde{m}_{Q}$}}
\def\msle{\mbox{$\tilde{m}_{L}$}}
\def\mqe{\mbox{$M_Q$}}
\def\mle{\mbox{$M_L$}}

\def\h#1#2#3{\mbox{$h_{#1#2}^#3$}}
\def\hd#1#2#3{\mbox{$h_{#1#2}^{#3^\dagger}$}}
\def\mh#1#2{\mbox{$h_{#1}^{H_#2}$}}
\def\mhd#1#2{\mbox{$h_{#1}^{{H_#2}^\dagger}$}}
\def\gam#1#2{\mbox{$\gamma_{#1}^{(#2)}$}}

\def\tw{\mbox{$\theta_{\tilde{w}}$}}
\def\tg{\mbox{$\theta_{\tilde{g}}$}}
\def\tsh{\mbox{$\theta_{\tilde{h}}$}}
\def\th{\mbox{$\theta_{H}$}}
\def\tq{\mbox{$\theta_{\tilde{q}_L}$}}
\def\tu{\mbox{$\theta_{\tilde{u}_R}$}}
\def\td{\mbox{$\theta_{\tilde{d}_R}$}}
\def\tl{\mbox{$\theta_{\tilde{l}_L}$}}
\def\te{\mbox{$\theta_{\tilde{e}_R}$}}
\def\tll{\mbox{$\theta_{L}$}}
\def\tqq{\mbox{$\theta_{Q}$}}
\def\h#1#2#3{\mbox{$h_{#1#2}^#3$}}
\def\hd#1#2#3{\mbox{$h_{#1#2}^{#3^\dagger}$}}
\def\mh#1#2{\mbox{$h_{#1}^{H_#2}$}}
\def\mhd#1#2{\mbox{$h_{#1}^{{H_#2}^\dagger}$}}
\def\gam#1#2{\mbox{$\gamma_{#1}^{#2}$}}
\def\gamg#1#2{\mbox{$\hat{\gamma}_{#1}^{#2}$}}

\newcommand\be{\begin{equation}}
\newcommand\ee{\end{equation}}
\newcommand\bea{\begin{eqnarray}}
\newcommand\eea{\end{eqnarray}}
\newcommand\beaa{\begin{eqnarray*}}
\newcommand\eeaa{\end{eqnarray*}}

\begin{document}

\begin{flushright}
IUHET-415
\end{flushright}

\begin{center}

{\Large{ Semi-perturbative unification with \\
extra vector-like families}}
\vskip .5in

{\large Mar Bastero-Gil$^{a}$ and Biswajoy Brahmachari$^{b}$}

\vskip 1cm

{\it (a) Department of Physics, University of Southampton,\\
 Southampton SO17 1BJW, UK \\
\vskip 1cm

(b) Physics Department, Indiana University, \\
Bloomington, IN 47405, USA \\
}

\vskip 0.5in

\end{center}

{%\singlespace

\begin{center}
\underbar{Abstract} \\
\end{center}

{
\small
We make a comprehensive analysis of an extended supersymmetric
model(ESSM) obtained by adding a pair of vector-like families to
the minimal supersymmetric standard model and having specific forms
of $5 \times 5$ fermion mass matrices. The singlet Higgs couplings
which link the ordinary to vector-like generations do not have the
renormalization effects of the gauge interactions and hence the
``quasi-infrared fixed point'' near the scale of the top quark mass. 
The two-loop Yukawa effects on gauge couplings lead to an unified 
coupling $\alpha_X$ around $0.2$ with an unification
scale $M_X$ of $10^{16.9}$ GeV. Large Yukawa effects in the high energy
region arrest the growth of the QCD coupling near $M_X$ making the
evolution flat. The renormalization effects of the vector-like 
generations on soft mass parameters has important effects on the 
charge and color breaking(CCB)minima. We will show that there exists 
parameter space where there is no charge and color breaking. We will 
also demonstrate that there exists minima of the Higgs potential which 
satisfies the mass of the Z boson but avoid CCB. Upper limits on the 
mass of the lightest Higgs boson from the one-loop effective scalar 
potential is obtained for sets of universal soft supersymmetry breaking 
mass parameters.
}

\newpage

\doublespace

\section{Introduction}

The commonly accepted notion of a unified coupling $\alpha_X=0.04$
at $10^{16}$ GeV has been questioned recently in view of a
conjecture that the four dimensional unified string coupling is
likely to have a semi-perturbative value (0.2-0.3) at the string
unification scale (see later) so that it may be large enough to
stabilize the dilaton but not as large as to ruin the
supersymmetric gauge coupling unification\cite{1}.

It may be noted that the MSSM unification scale
$M_X=2 \times 10^{16}$ GeV appears to be a factor of 20 smaller
than the one-loop level string unification scale of $M_{st} \sim
g_{st}~ 5.2 \times 10^{17}~{\rm GeV} \sim 3.6 \times 10^{17}$
GeV\cite{2,3}. In this context the extensions of MSSM spectrum 
addressing the issues of $\alpha_X$ and $M_X$ may be interesting. The
specific extension which we will consider is called as
the Extended Supersymmetric Standard Model (ESSM). This itinerary 
extends the MSSM particle spectrum by adding two light vector-like 
families which are $Q_{L,R}= (U,D,N,E)_{L,R}$ and $Q^\prime_{L,R}=(U^\prime,
D^\prime,N^\prime,E^\prime)_{L,R}$ and two Higgs singlet 
scalars $H_S$ and $H_\lambda$. These new particles and
as well as their superpartners are 1 to a few TeVs\cite{1}. ESSM has
no unified gauge group as such and so the corresponding 
leptoquark and di-quark gauge bosons leading to proton decay
does not exist. The extra sets of matter fields denoted by previous
authors $Q_L|{\overline Q^\prime_R}$ and $\overline{Q_R} | Q^\prime_L$
can be thought of as ${ 16}$ and ${ \overline{16}}$ representations 
of a cosmetic unified gauge group of SO(10).

It has been noted sometime ago that precision measurements of the
oblique electroweak parameters and the number of light neutrino
species $N_\nu$ does not favor extra chiral families. 
They do not however constrain vector-like families due to decoupling
effects\cite{5,peskin}. Such vector-like families i.e., ${16}$ 
and ${ \overline{16}}$, which apparently are predicted
generically in string theories could well exist in the TeV region.
It may be worth understanding that the pair of vector-like
families could give us a clue to the family mass hierarchy within
the three chiral generations via a see-saw mechanism residing
inside the $5 \times 5$ fermion mass matrices\cite{6}.

Such a  pair of vector-like families and thus ESSM provide some unique
advantages over MSSM. They are as follows.

\noindent $\bullet$  One-loop evolution of three gauge couplings
maintains the approximate meeting when complete extra 
families such as vector-like families are included.

\noindent $\bullet$ Although one-loop approximation leads to
the unification of the three $\alpha_i$'s at the same
scale $M_X$ the unified coupling $\alpha_X$ increases.

\noindent $\bullet$ Numerically speaking $\alpha_X$ can be raised
to $0.2-0.3$ in one loop. Two-loop gauge couplings accelerate the
growth whereas the two loop Yukawa effects on gauge couplings does
the contrary. This is vital for having a good fit to $\alpha_s(m_Z)$
within unification.

Now we briefly describe the work of the previous authors\cite{1}. Instead
of using the $\theta$ function approximation to study the particle
thresholds they used exact threshold functions \cite{kennedy,7}. The
evolution of the Yukawa couplings was studied in one-loop. The
off-diagonal Yukawa couplings between the third family and the
vector-like families and the self couplings of the vector-like
families were taken to be large varying between 1.0-2.0. This was done for
the Yukawa couplings of the up, down, charged lepton and neutrinos so that
they approach their ``quasi infrared fixed point'' values near the
scale of the top quark mass. The combined effects of the two-loop gauge
contributions with one-loop Yukawa contributions and threshold effects
exhibited the following features.

\noindent $\bullet$ Three couplings still met. This was for a wide range
of variation of the particle spectrum. A higher $\alpha_X \sim 0.2-0.3$
as well as higher $M_X \sim 10^{17}$ GeV compared to the MSSM
values $\alpha_X \sim 0.04 ~~{\rm and}~~ M_X \sim 2 \times 10^{16}$ GeV
was found. ESSM unification scale $M_X$ is now closer to the
string scale. The remaining gap of a factor of 5 or so between $M_X$ and
$M_{st}$ may not be alarming. It could partially be due to the increased
influence of the two-loop string threshold effects. This is because a
semi-perturbative range of values of $\alpha_X$ is at work. Such a value
can bring a correction to the one-loop formula of $M_{st}$.

\noindent $\bullet$ The couplings meet for lower values of
${\alpha_3(m_Z)|_{\overline{MS}}}^{ESSM} \sim 0.120-0.124$ in the
ESSM case. This is in good agreement with the experimentally observed
value of $\alpha_3(m_Z)|^{obs}_{MS} = 0.117 + \pm 0.005$. By contrast 
we typically need higher values of $\alpha_3(m_Z)$ such as
$\alpha_3(m_Z)|_{\overline{MS}}^{MSSM} > 0.125$ if $m_{\tilde{q}} < 1 $
TeV and $M_{1/2} < 500 $ GeV in the MSSM case \cite{7}.

These promising features serve the motivation to study
ESSM further. The purpose of this paper is to explore the unification
of ESSM gauge couplings further by including two-loop Yukawa
evolution(see figure 1.b). Three-loop evolution of the three gauge
couplings including vector-like families with contributions only from
gauge interactions has already been studied to some extent\cite{8}. Our
reason to study the two-loop evolution of the Yukawa couplings of
ESSM is that there are 20 Yukawa coupling entries. These
couplings can be near the perturbative maximum
\footnote
{
This may be
viewed in comparison to the non-perturbative values of the unified
couplings \cite{9}.
} 
at $M_X$. We also expect that the two-loop
contributions of the semi-perturbative gauge couplings to
the evolution of the Yukawa couplings would be
significant
\footnote
{
We would estimate the ratio of
two-loop versus one-loop contributions to the $\beta$ functions
for the Yukawa couplings to be $ \sim [(g^2 {\it or} h^2) / 4 \pi]
(4 \pi)^{-1}$ C, where the coefficients C are typically large.
Taking a representative value of $C \sim 10-30$ it suggests that
the ratio of relative contributions can be about 25 to 75 \% near
$M_X$ for $g^2/4 \pi \sim h^2/4 \pi \sim 0.25$.
} 
\cite{10}.

It seems to us that it is important to ask the question whether the
inclusion of the two-loop contributions of the gauge couplings to the
evolution of the Yukawa couplings and  vice-versa would preserve, improve
or adversely alter the features of ``precision'' coupling unification
studied previously\cite{1}. Thus we think that two-loop gauge evolution
coupled to two-loop Yukawa evolution is an interesting approximation
to explore in the context of semi-perturbative unification.

\section{Evolution of couplings at two-loops}

\subsection{Definitions}

We perform a two-loop analysis of the renormalization group
equations (RGE) of the gauge couplings including the effects of
the third generation Yukawa couplings at two-loops and carefully
taking into account the threshold effects due to the spread of the
superpartner masses and the masses of the particles of the
vector-like families. Furthermore we use mass dependent running of
the gauge couplings as given in(\ref{ale1}) or equivalently stated
as the effective coupling formalism which guarantees a smooth
cross over of the beta function coefficients at the threshold of each
individual particle. The two-loop RGEs for the Yukawa couplings can be
found in the Appendix and those for the gauge couplings can be expressed as
\be
\frac{d\alpha_i^{-1}}{dt}= -\frac{b_i}{2 \pi} -\sum_j\frac{b_{ij}}{8 \pi^2}
\alpha_j+\sum_k\frac{a_i^{(k)}}{8 \pi^2} Y_k\,.
\label{rge2}
\ee
We have redefined $t=\ln \mu$ and $Y_k=h_k^2/4 \pi$. The coefficients
$b_i$, $b_{ij}$ and $a^{k}_i$ are given in the Appendix. In
a mass dependent subtraction procedure these coefficients are
replaced by threshold functions\cite{kennedy}. The threshold
functions depend
on the scale $\mu$ through the ratios $m_k/\mu$ where $m_k$ is the
mass of the particle running inside the loop. Integrating
(\ref{rge2}) we obtain the analytical expression of the effective
couplings. At one-loop approximation they are given by
\be
\alpha_i^{-1}(\mu)=\alpha_i^{-1}(\mz)+ \sum_k \frac{b_i^k}{2 \pi} \left(
F_k(m_k/\mu)- F_k(m_k/\mz) \right) \,.
\label{ale1}
\ee
The threshold function $F_k(m_k/\mu)$ has the value 
$\ln (m_k/\mu)$ in the limit
$m_k/\mu \rightarrow 0$. Therefore when all the masses are
well below the scale $\mu$ we recover the familiar one-loop expression
\begin{equation}
\alpha_i^{-1}(\mu)=\alpha_i^{-1}(\mz) -{b_i \over 2 \pi}
\ln\left( {\mu \over m_Z }\right).
\end{equation}
On the other hand $F_k(m_k/\mu)$ vanishes in the limit $m_k/\mu \rightarrow
\infty$ showing the familiar decoupling of heavy masses from the running
of the couplings. A smooth threshold crossing with the variation
of $\mu$ is obtained thereby. In our approximation the two-loop parts of
the beta functions $b_{ij}$ and $a_i^{k}$ changes at each threshold in a
step-function approximation though. The details of our procedure to
determine the masses of  the superpartners with the constraint of correct
radiative electroweak breaking in ESSM will be given in the following
sections.

At the unification scale $M_X$ we have the Yukawa coupling matrix of the
third and
the heavy vector-like generations ${ 16}$
and ${ \overline{16}}$ in the simple form 
\begin{eqnarray}
h_{f,c}^{(o)} =
\bordermatrix
{& 16_3 & { 16} & \overline{{ 16}} \cr
   16_3 & 0 & x_f H_{f} & y_f H_{\rm S}\cr
   { 16}   & y^{\prime}_c H_{S} & 0 & z_f H_{\lambda} \cr
\overline{{ 16}} & x^{\prime}_c H_{f} & z^\prime_c H_{\lambda} &
0\cr}~.
\label{mssmyuk}
\end{eqnarray}
The couplings $x_f$, $x^\prime_c$ ($y_f$, $y^\prime_c$) denote the
interactions between the chiral and the vector-like generations
via the doublet scalars $H_1$, $H_2$ and singlet scalar $H_S$
whereas  $z_f$, $z^\prime_c$ give the interactions within the
vector-like generations through the singlet $H_{\lambda}$. Note
that this is only a formal representation of the Yukawa matrix. 
In practice they should be viewed in terms of the component 
Yukawa matrices given in the Appendix. When we run down the
couplings from the unification scale to the electroweak scale
the vector-like generations become massive and consequently the
Yukawa matrices get rotated projecting the effective third
generation Yukawa couplings $h_t$, $h_b$ and $h_{\tau}$ at \mz.
Hence the RGE for the Yukawa couplings are integrated in a
top-down approach in the ranges $M_X \rightarrow M_Q \rightarrow
M_L \rightarrow \mz$, with the boundary conditions at the
vector-like scale  \bea
  h_t(M_Q)&=& \left. (\frac{x^\prime_u y_u}{z_u} + \frac{x_u y^\prime_q}{z^\prime_q})
\frac{v_S}{v_{\lambda}}\right|_{M_Q}\,, \nonumber \\
  h_b(M_Q)&=& \left. (\frac{x^\prime_d y_d}{z_d} + \frac{x_d y^\prime_q}{z^\prime_q})
\frac{v_S}{v_{\lambda}}\right|_{M_Q}\,, \nonumber \\
  h_{\tau}(M_L)&=& \left. (\frac{x^\prime_e y_e}{z_e} + \frac{x_e y^\prime_l}{z^\prime_l})
\frac{v_S}{v_{\lambda}}\right|_{M_L}\,.  \label{bounc}
\eea
Here $M_Q$ denote the quark and $M_L$ denote the
leptonic members of the extra generations
consequently. Furthermore we call $v_S=<H_S>$ and $v_\lambda=<H_\lambda>$.
Assuming all
the Yukawa couplings to be large at the
unification scale ($h_i(M_X)= \sqrt{4 \pi}$) we evolve them to the scale
$m_Z$. For the sake of a compact notation let us give it the
name complete fixed point scenario(CFPS). We fix $v_S/v_\lambda
\sim 0.5$ for the time being. This prevents a low value
of $h_t(\mz)$ in terms of the ESSM couplings described in (\ref{bounc})
and consequently prevents a low prediction of the top quark mass. 
Later we will
prove that there exists charge and color conserving
minima with our choice of $v_S/v_\lambda \sim 0.5$. Furthermore we see
that the singlet Higgs couplings do not feel the QCD
interaction and thus cannot approach a ``quasi-infrared fixed point'' near
the scale of the top quark mass and thus we may suspect that the radiative
electroweak breaking may occur at the wrong place if their initial
value at $M_X$ is taken at $\sqrt{4 \pi}$.

Imposing the unification condition 
$\alpha_1(M_X)=\alpha_2(M_X)=\alpha_3(M_X)=\alpha_X$ the
gauge couplings are evolved down to \mz~ scale. We use the central values
of $\sin^2~\theta_w=0.2319$ and
$\alpha^{-1}_{em}(\mz)=127.9$ to fit $M_X$ and $\alpha_X$. 
Now $M_X$ and $\alpha_X$ are fitted (which depend on the mass spectrum
of the vector-like generations and superpartners as well as
Yukawa couplings) so we predict the QCD coupling $\alpha_3(\mz)$ by
running back using two-loop RGE. 

To compare the values of $\alpha_i(\mz)$ with the
experimental ones we have consistently translated the effective
couplings to the values in the \mins\ scheme. We have
found that in the present case  $U(1)$
and $SU(2)$ effective couplings are only 1\% larger than the ones
in the \mins\ scheme. However $\alpha_3(\mz)$ can be 5--8\% larger
depending on the supersymmetry spectrum under consideration.

\subsection{Comparison of one and two-loop results}

\begin{figure}[t] 
\begin{tabular}{cc} 
\epsfysize=7cm \epsfxsize=7cm \epsfbox{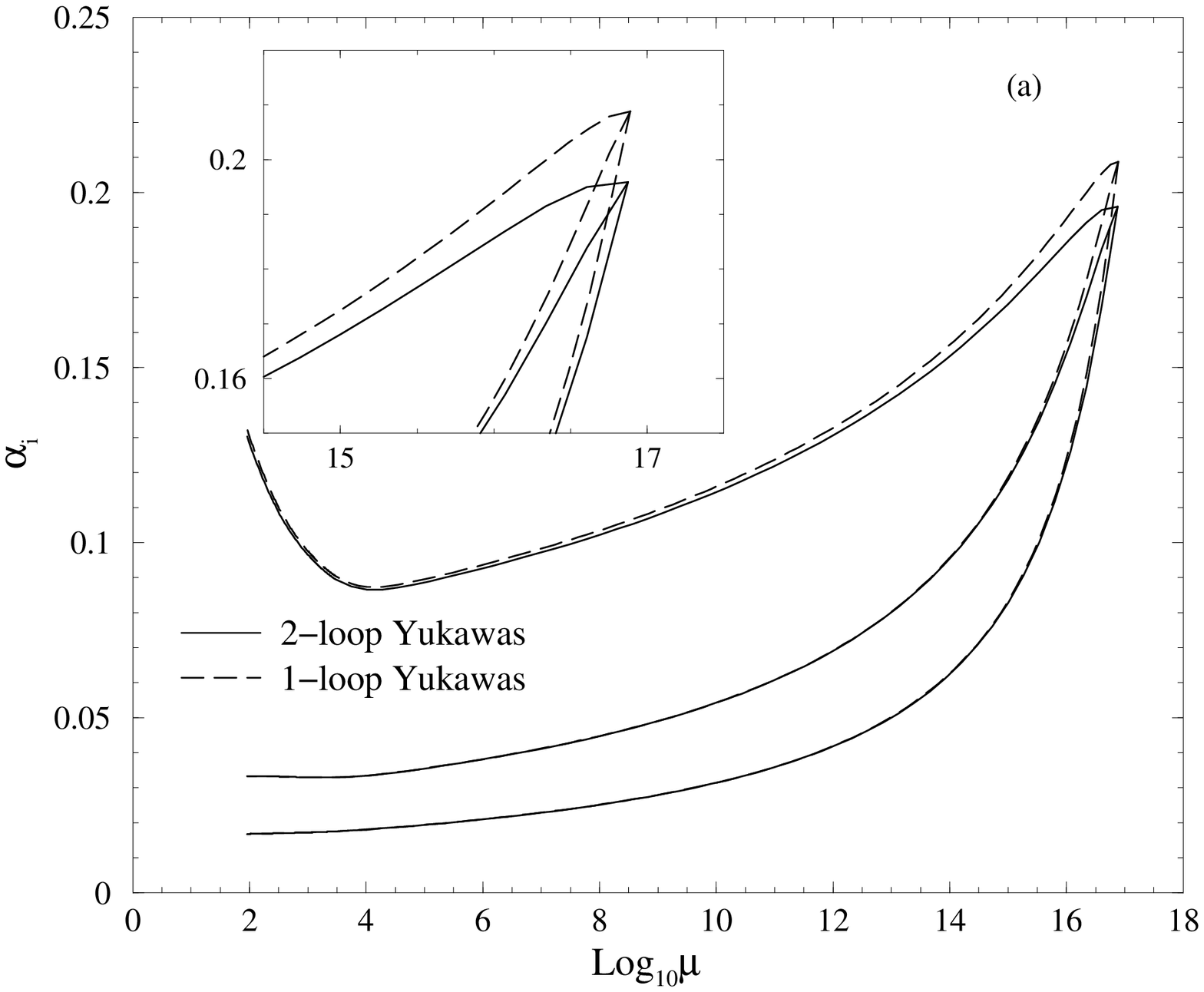}  & 
\epsfysize=7cm \epsfxsize=7cm \epsfbox{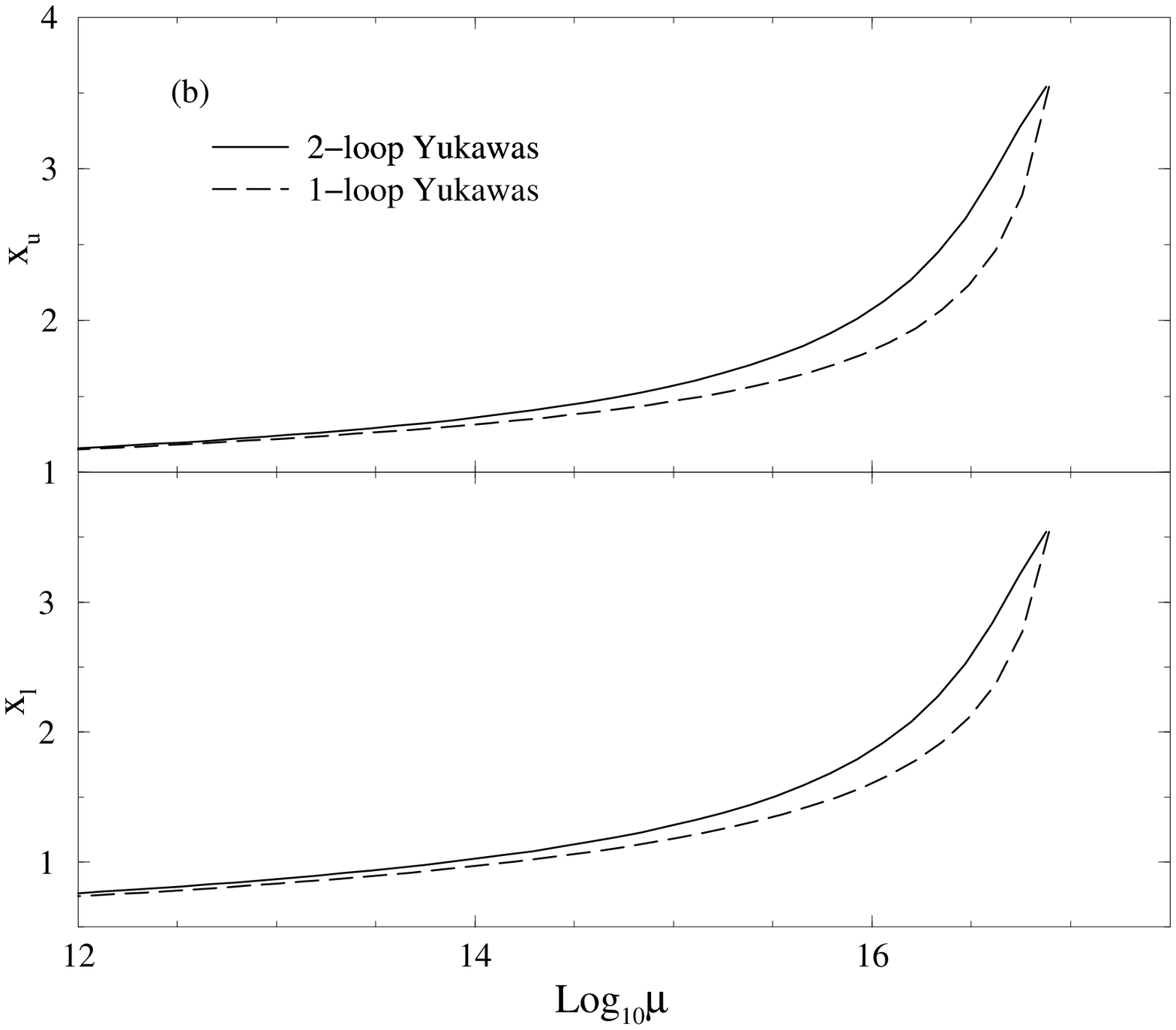}  \\
\end{tabular}
\caption{(1.a) Running of the gauge couplings when the Yukawa
couplings are taken at one loop (dash line) and at two-loop (solid
line). (1.b) Running of the Yukawa couplings at one-loop (dash
line) and two-loops (solid line). $M_2=90\, $ GeV,
$\tilde{m}_{q}=\tilde{m}_h=1\, $ TeV, $M_Q=3 \, $ TeV.\label{ali}}
\end{figure}

Let us choose a feasible superparticle 
mass spectrum which can be used to evaluate the threshold functions. 
Later we will calculate the evolution of the
soft mass spectrum more accurately using universality and use iterative
numerical subroutines to calculate the threshold functions.
We choose $M_Q=3$ TeV, wino mass $M_2=90\, $
GeV, squark mass \msq\  and higgsino mass \msh\ as 1 TeV. The
more massive Higgs scalars are also taken at 1 TeV while keeping
the lightest Higgs mass at 95 GeV. We have neglected the mixing between
the charginos and
higgsinos as well as between left and right handed squarks and
sleptons which occur in the threshold functions. We have assumed the mass
pattern of the vector-like
generations as $\mqe \sim \msqe$ and $\mle \sim \msle$. 
This reflects the assumption that supersymmetry breaking scale is 
near 1 TeV whereas the vector-like scale is few TeVs. Thus we are led 
to the approximate relation $\mqe \sim 3 \mle$. The factor of 3 within 
the vector-like generations represent the similar QCD
renormalization effects parallel to the three chiral generations.
In figure \ref{ali}.a we have compared the evolution of the effective
gauge couplings coupled with the Yukawa couplings. Yukawa couplings  are
calculated at one-loop(dash line) and two-loops(solid line).

We first notice that in the high energy region the two-loop
Yukawa couplings are appreciably different from the one-loop ones
(figure \ref{ali}.b). The curves for the
gauge couplings are flattened near the
unification scale. Note that effect is pronounced in the case of
the QCD coupling. Due to the presence of a
large number of Yukawa couplings their
contributions to gauge coupling beta function coefficients
becomes comparable the gauge coupling contributions. They may even be
mutually canceling causing a flatness of the curve. This cancellation 
is more forceful when the Yukawa couplings are integrated at 
two-loops. At low energy the Yukawa couplings
approach their ``quasi-infrared fixed points''. The
distinction between the one-loop and the two-loop results for the
Yukawa couplings reduce in the low energy region below the
vector-like threshold. From the point of view of gauge coupling
unification the overall effect would be a small increase of the
values of $\alpha^{-1}_X$ and $\alpha^{-1}_3(m_Z)$ and $M_X$.

Stage is set for the question, ``how much these predictions vary when the
scale $M_Q$ is varied ?'' Intuitively it is clear that we should recover
the MSSM results in the limit $M_Q \rightarrow M_X$. In figure
\ref{alQ} the predictions of $\alpha_X$ and $\alpha_3(m_Z)$ are
plotted when varying $M_Q$ with Yukawa couplings at one-loop
(curves labelled (a)) and two-loop Yukawa couplings (curves
labeled (b)). They also describe the results including the
threshold effects due to a higher gaugino
mass of $M_2=200$ GeV. Once $\tilde{m}_{q_L}$ and $M_2$ are fixed
so are $M_0$ and $M_{1/2}$ at the unification scale (figure
\ref{m0m12}).

\begin{figure}[t]
\begin{tabular}{cc} \epsfysize=7cm \epsfxsize=7cm 
\epsfbox{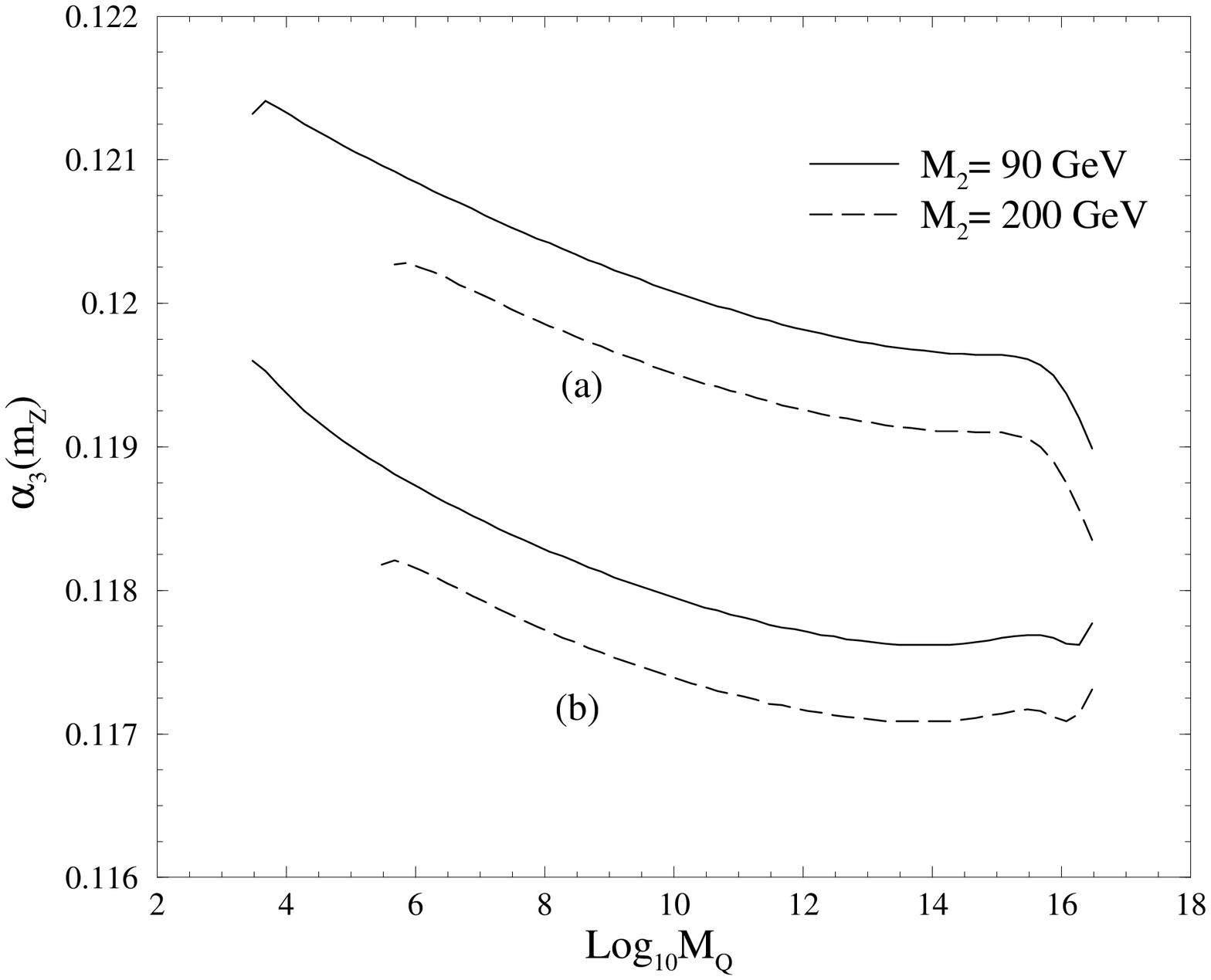}  & \epsfysize=7cm \epsfxsize=7cm 
\epsfbox{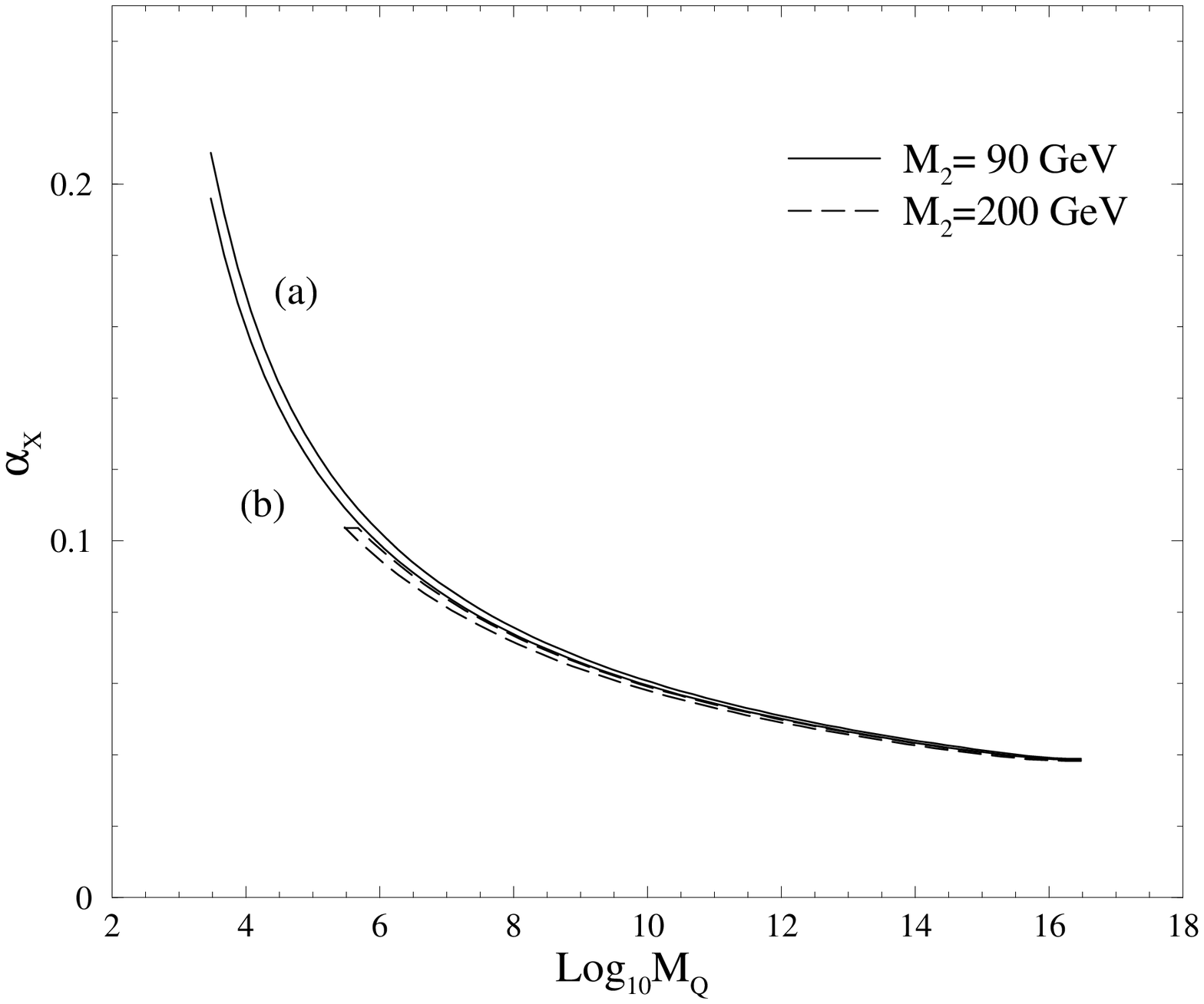}  \\
\end{tabular}
\caption{ The predicted values of $\alpha_3(m_Z)$ and $\alpha_X$
with varying $M_Q$. The solid and dash lines are for different
$M_2$ as given in the legend. Curves labelled (a) are obtained
with the Yukawa couplings  integrated at one-loop whereas those
labelled (b) shows the Yukawa couplings at two-loops.
\label{alQ}}
\end{figure}

The plotted values of $\alpha_3(\mz)$ are their \mins\ scheme
value. When $M_Q$ is near the TeV  range 
the three gauge couplings are
non-asymptotically free for most of the range of integration and later
they
merge at a larger value of $\alpha_X$ compared to the MSSM
prediction. When $M_Q$ increases asymptotic non-free  behavior
reduces and the MSSM results are recovered.

\subsection{Fixed point scenario, vector-like scale and fermion masses}

Now we have to choose an input value of the ratio $v_S/v_\lambda$ to
translate Yukawa couplings to fermion masses using (\ref{bounc}). We
continue with our choice of the ratio $v_S/v_\lambda \sim 0.5$. Next we
have to fix all Yukawa couplings of isospin
up sector ($H_2$ coupling) as well as isospin down
sector($H_1$ coupling) at $M_X$. This is done by choosing both sets
at $\sqrt{4 \pi}$. After evolving the Yukawa couplings the related 
fermion
mass scales we input $m_\tau =h_\tau~\cos \beta~{V_F \over \sqrt{2}}=
1.777 \, $ GeV for
the isospin down sector and calculate
the familiar  $\tan \beta$ which is $<H_2> \over <H_1>$
of MSSM. At this stage we
can check the predictions of the top quark mass and the bottom quark mass.
We have  found $m_t^{pole}= 165 -
180 \, $ GeV, $m_b(m_b) = 4.1 - 4.4 \, $ GeV. The range corresponds to
a wide variation in $M_Q$ is in the domain $3$ TeV-$10^{16} \, $ GeV.
Each value of $M_Q$ gives a value of $\tan \beta$ via $m_\tau$.
The range of $\tan \beta$ 35 to 60 is simply reflecting that all the
Yukawa couplings at $M_X$ are $\sqrt{4 \pi}$.

\begin{figure}[t]
\begin{tabular}{cc} 
\epsfysize=7cm \epsfxsize=7cm \epsfbox{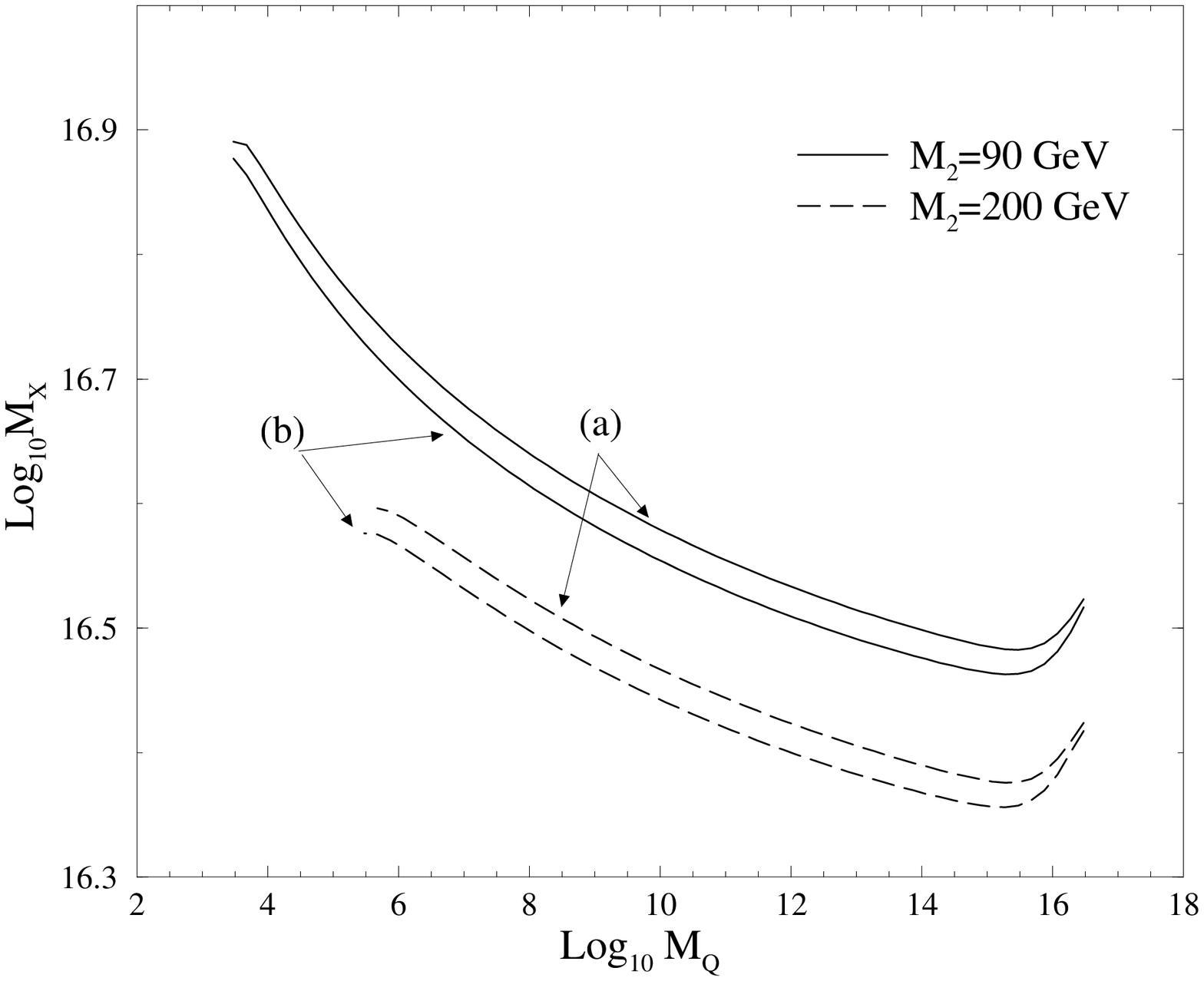}  & 
\epsfysize=7cm \epsfxsize=7cm \epsfbox{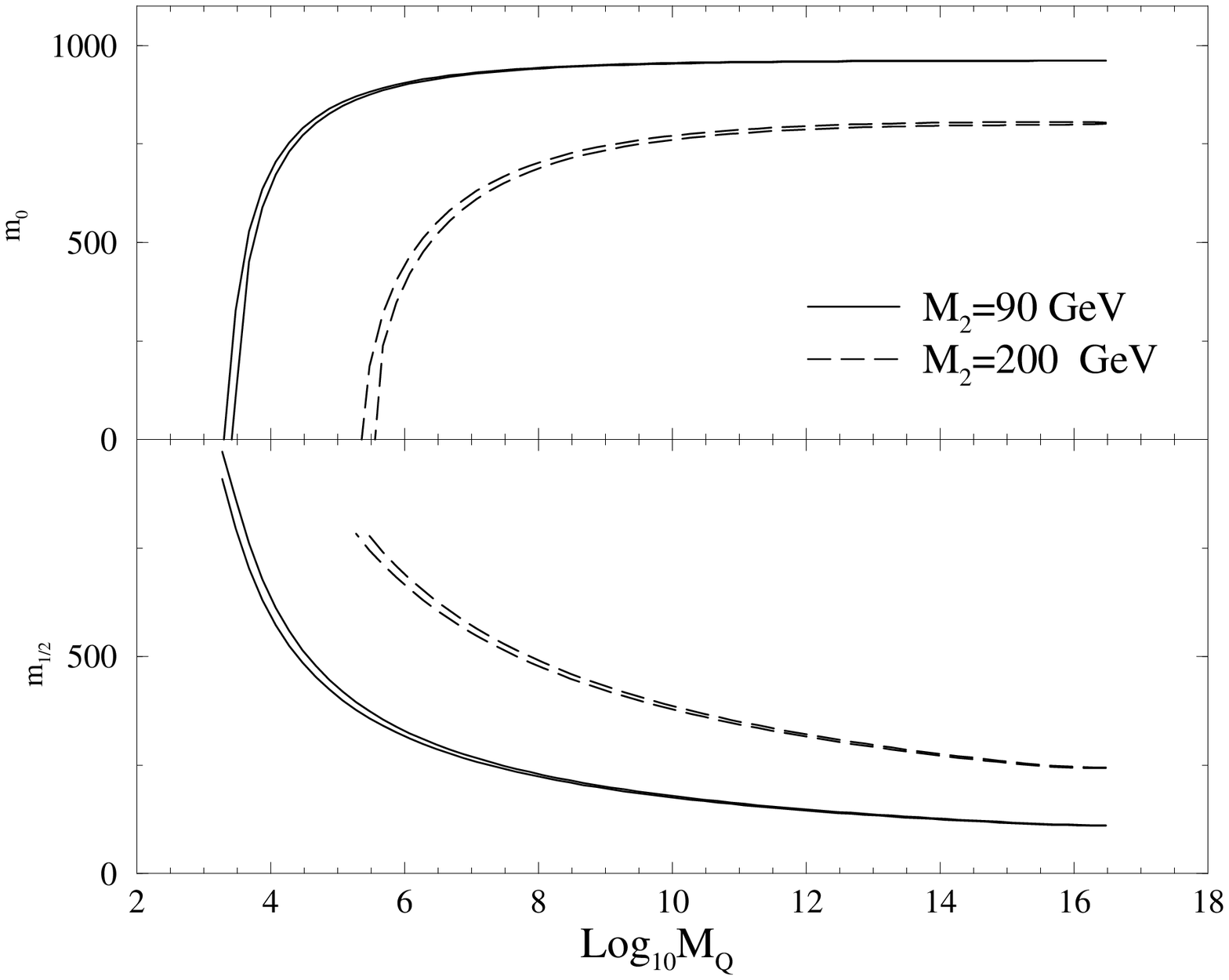}  \\
\end{tabular}
\caption{The figure at left gives the predicted values of $M_X$; labels
(a) and (b) imply Yukawa couplings integrated at one-loop or two-loops.
The figure in the right gives the extrapolated values of
$M_0$ and $M_{1/2}$ (at the unification scale)
requiring that the lightest charging mass $M_{1/2}$ is
90 GeV. First and second generation squark masses are consistently
taken at 1 TeV.
\label{m0m12}}
\end{figure}

An interesting experimental situation from the point of view
of semi-perturbative unification would be to have the  vector-like
scale of a few TeVs. The mass of the
vector-like particles can be related to an ``effective'' $\mu$
parameter. We will define it in a moment. A  simple  numerical
estimation will show us that if we demand that the vector-like squark
masses
are at the TeV region the scenario with all the
ESSM Yukawa couplings are $\sqrt{4 \pi}$ at $M_X$ is
excluded. It will be apparent that the reason is related to the link
between $\mu$ and $v_\lambda$.

The superpotential for the Higgs scalars is
\begin{equation}
W= k_1 H_1 H_2 H_{\lambda} + \frac{k_2}{3} H_{\lambda}^3 +
\frac{k_3}{3} H_S^3.
\end{equation}
This specific superpotential
is an inspired guess as suggested in\cite{1}.
The scalar $H_S$ plays little role in the EWSB as it does not
couple to $H_1$ and $H_2$.
However the singlet
$H_{\lambda}$ plays the role of the field $N$ in the Next-to-Minimal
Supersymmetric Standard Model
(NMSSM)\cite{fayet,frere,ellis}, and EWSB is driven when $N$ gets a
VEV (see section 3.2 and figure 4 later). An effective $\mu$  parameter
can be defined by the relation
\begin{equation}
\mu \equiv k_1 v_{\lambda} \,.
\end{equation}
The coupling $k_1$ is computed near $m_Z$ scale. We
expect that $\mu \sim 1\, $ TeV. The vector-like fermion masses
are obtained as the eigenvalues of the matrix given in
(\ref{mssmyuk}). In the first approximation they are
\begin{equation}
M_{U1} \sim M_{D1} \sim z^\prime_q(M_Q) v_\lambda,\;\;\;M_{U2}\sim
z^\prime_u(M_Q) v_\lambda,\;\;\;M_{D2} \sim z^\prime_d(M_Q) v_\lambda,
\end{equation}
and similarly for the heavy leptons, where the ratio
$z^\prime_q(M_Q)/z^\prime_l(M_L)\sim 3 $ is found due to
renormalization effects. It is very easy to see that 
we have the ``quasi-infrared
fixed points'' after integrating the RGE
\footnote
{
$k_1$ does not feel QCD interaction however it still feels the SU(2)
interaction via $H_1$ and $H_2$ which cannot pull it up too much when the
scale approaches $m_Z$ scale from above.
}
\begin{equation}
z^\prime_q(M_Q) \sim 0.73, \;\;\; z_u(M_Q) \sim 0.56, \;\;\;
z_D(M_Q) \sim 0.53,~~{\rm and}~~k_1(M_Q) \sim 0.01. \label{fp}
\end{equation}
We see $v_\lambda=
\mu/k_1 \sim 100\,{\rm TeV}$ and thus the vector-like particles
are at a scale $\sim 50
\,{\rm TeV}$. This is far too large to be interesting. Therefore if we
consider smaller values of the vector-like
scale the initial values of the couplings $z_f$ and $z^\prime_c$
at the unification scale have to be smaller so as to fit the input value
of $M_Q$ of few TeVs.  
This is our first reason to reduce the Yukawa couplings from the
CFPS scenario. As a matter of fact smaller Yukawa
couplings are forced by potential minimization conditions
for any value of $M_Q$ as described later.

In the next section we will study the soft mass evolution. But
beforehand we may take note that the effect of large
Yukawa couplings is to diminish the superparticle masses.
Large Yukawa couplings will dictate  charge and color breaking
minima driven by cubic $A$ couplings even when the { weak} 
constraint\cite{frere,alvarez}
\begin{equation}
|A_{ijk}|^2 < 3 ( m_i^2 + m_j^2 + m_k^2 ) \label{aijk},
\end{equation}
is satisfied.
We must stress that
especially when all the Yukawa couplings are at the
``quasi-infrared fixed point'' region near the scale of the top
quark mass a detailed evaluation of the minimization conditions 
are necessary\cite{casas}. We will
demonstrate  in this case there is a charge and color breaking
minimum deeper than the electroweak symmetry breaking one in ESSM
for any value of $M_Q$ including the MSSM limit. Consequently we will
conclude that one or more Yukawa coupling must be reduced at the
unification point to reach a {global} charge and color conserving minimum.

\section{Soft mass spectrum of ESSM}

\subsection{Soft mass evolution}
Assuming the universality of soft supersymmetry breaking terms at the
unification scale the supersymmetry spectrum at the weak scale is given
in terms of following parameters at the unification scale apart from 
the dimensionless gauge and Yukawa couplings 
\be
{\rm gaugino~mass}=\m12 ~~;~~{\rm scalar~mass} =\ms0~~;~~{\rm cubic~
coupling}= A_0.
\ee
The gaugino masses $M_i$ are calculated integrating the two-loop RGE
\be
\frac{dM_i/\alpha_i}{dt} = \sum_j\frac{b_{ij}}{8 \pi^2} \alpha_j^2
\frac{M_j}{\alpha_j}.
\ee
The squark and slepton masses of the first and second generations are
calculated neglecting small D-term contributions from the one-loop
expression of
\begin{equation}
\frac{d \tilde{m}_a^2}{d \ln \mu} = - \frac{1}{2 \pi} \sum_i 4 C_2(R_a)
\alpha_i M_i^2 \,.
\label{1msa}
\end{equation}
We have $C_2(R_a)=(N^2-1)/2 N$ for the fundamental representation of
$SU(N)$. When calculating the squark and slepton masses of the third
generation we have properly taken into account the effects of the related
Yukawa couplings. Here we take $\sqrt{4 \pi}$ as the initial
value at the unification scale for those  Yukawa couplings of ESSM
which are related to $h_t(m_t)$ below the vector-like threshold.
This fixes the top quark mass.

The one-loop RGE as in (\ref{1msa}) can be easily integrated numerically.
It is convenient to represent them after integration by the expression
\begin{equation}
\tilde{m}_a^2= \mss0+ c_a \mww \,.
\label{maca}
\end{equation}
The constants $c_a$ depend on the values of the gauge
couplings and their $\beta$-functions and thus implicitly(via threshold
effects) on the
vector-like scale $M_Q$.
We again rewrite (\ref{maca}) in terms
of the wino mass $M_2$ instead of the universal mass parameter $M_{1/2}$
in the following expression
\begin{equation}
\tilde{m}_a^2= \mss0+ \hat{c}_a M_2 \,.
\label{maca12}
\end{equation}
The values of $\hat{c}_a$ are much larger than those of MSSM.
Sample values of $\hat{c}_a$ are given in table \ref{table1}. The
MSSM limit is $M_Q \sim 10^{16.5}$ GeV is also
quoted in table \ref{table1} for instant comparison.
ESSM spectra gives an 
enhancement of the squark and slepton masses for a given $M_{1/2}$
compared to MSSM. This was noticed previously\cite{8}.

Table 1 gives such coefficients for the masses of the left handed
squarks
\footnote
{
Those for the right handed squarks, up and down,
are roughly the same as $\hat{c}_Q$. In the numerical
calculations of the effective couplings we have assumed degenerate
masses of left and right handed squarks where both are taken to be
$\tilde{m}_q$.
}
, left and right handed sleptons. We
also give the coefficient for the ratios of
the gaugino masses. Here we have  taken the initial
values Yukawa couplings at the largest allowed value (see
illustration after (\ref{fp})). Of course later on
we will fix the Yukawa couplings via potential minimization. For
low values of $M_Q$  the squark and slepton masses of the first
and second generations get large gaugino contribution due to the
absence of the reverse effects from the Yukawa couplings. Hence
even if we assume that the wino is as light as the experimental
lower limit pushing  $M_{1/2}$ to the least, the squark masses would
still move to the TeV range . The left-handed slepton mass is larger 
than the gluino mass $M_3$ and the right-handed slepton mass is roughly 
of the same magnitude as $M_3$. As $M_Q$ increases we reach the MSSM 
limit where light gaugino masses together with
large squark masses imply a large value of $M_0$ as can be seen in
figure \ref{m0m12}. It is interesting to observe that the ratio
${\msg \over \msw} \sim 4$ is practically independent of $M_Q$.
\begin{table}
\begin{center}
\[
\begin{array}{|c|c|c|c||c|c|c|}
\hline
M_{Q}& \hat{c}_Q& \hat{c}_L& \hat{c}_R& M_{1/2}/M_2& M_3/M_2& M_2/M_1\\
\hline
3\, TeV        & 110.2 & 21.2 & 8.8 & 8.8 & 3.8  & 1.6 \\
300\, TeV      &  27.3 &  4.7 & 1.8 & 4.0 & 3.9  & 1.8 \\
10^{10}\, GeV  &  10.7 &  1.2 & 0.4 & 1.9 & 4.0  & 1.9 \\
10^{16.5}\,GeV &   9.3 &  0.8 & 0.2 & 1.2 & 4.0  & 1.9 \\
\hline
MSSM           &   9.3 &  0.8 & 0.2 & 1.2 & 4.0  & 1.9 \\
\hline
\end{array}
\]
\end{center}
\caption{Coefficients $\hat{c}_a$ for the left-handed squarks, left
and right handed sleptons masses of the first and second generation. The
MSSM limit is at $M_Q \sim 10^{16.5}$ GeV.}
\label{table1}
\end{table}

\subsection{Radiative electroweak breaking}

The Higgs sector of the ESSM resembles to that of the
Next-to-Minimal Supersymmetric Standard Model (NMSSM).
The VEV of $H_\lambda$ generates  effective $\mu$ and $B$
parameters. In other words at low energy we evaluate the
expressions
\begin{eqnarray}
\mu &\equiv& k_1 v_{\lambda} \,, \label{mu}\\
B &\equiv& k_2 v_{\lambda} + A_1 \label{b} \,.
\end{eqnarray}
$A_1$ and $A_2$ are the cubic dimension-full couplings corresponding
to the dimensionless Yukawa couplings $k_1$ and $k_2$ of the
superpotential {\it W}.

In MSSM there are two minimization conditions. They are the
equations obtained by minimizing the scalar potential with respect to
the real components of $H_1$ and $H_2$. Using these conditions
one can fit two parameters $\mu$ and $B $ with two inputs $\tan \beta$
and $m_Z$\cite{mar}. In ESSM there are three minimization 
conditions. The conditions for the minimization with respect to 
the real components of $H_1$ and $H_2$ along with the extra condition  
\begin{equation}
\sin 2 \beta= 2 v_\lambda\frac{ m_{H_\lambda}^2 + k_2 A_2 v_\lambda +
k_1^2 v^2 + 2 k_2^2 v^2_\lambda}{k_1 A_1 v^2 + 2 k_1 k_2 v^2
v_\lambda} \,.
\end{equation}
These three conditions can be used to fit three parameters
($k_2, A_0, v_\lambda$) with three inputs ($\tan \beta, k_1(M_X), m_Z$).
We have taken $k_1(M_X)=\sqrt{4 \pi}$. Once $A_0$ is fixed the 
effective $\mu$ and $B$ parameters were computed
using (\ref{mu}) and (\ref{b}) for various values of $M_0$ keeping
$M_2=90$ GeV. The values are displayed in figure \ref{mub0}. The value 
of $\mu$ is roughly constant with the variation of $\tan \beta$ at the 
TeV scale. The $B$ parameter decreases with $\tan\beta$. 

\begin{figure}[t]
\epsfysize=8cm \epsfxsize=8cm  \epsfbox{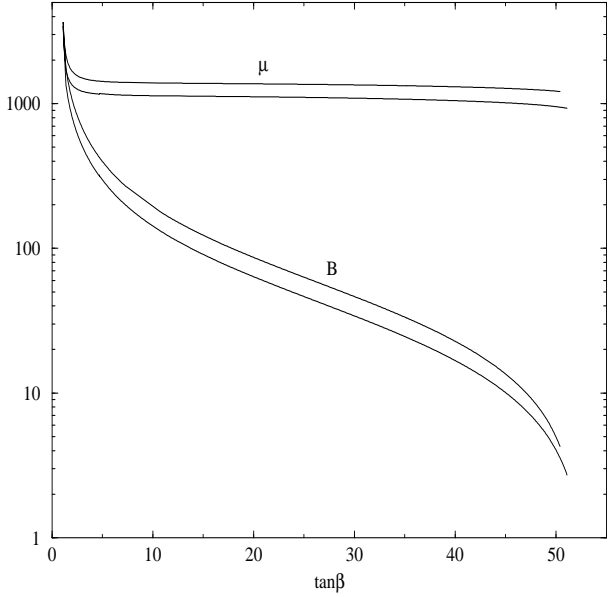} 
\caption{Fitted values of the effective parameters $\mu\equiv k_1(m_Z)
v_\lambda$ and $B\equiv k_2(m_Z) v_\lambda+A_1$ at Low energy. For each
parameter the lower curve correspond to $M_0=0$ and the upper curve to
$M_0=800$ GeV. We have taken $k^2_1(M_X)={4 \pi}$.
\label{mub0}}
\end{figure}

\subsection{Superparticle masses of ESSM}

Let us remind ourselves that while calculating the unification coupling
and unification scale (in the figures \ref{alQ} and \ref{m0m12} ) we
did not take into account the constraint from electroweak symmetry
breaking (EWSB). We showed how the predictions depend on
the  vector-like scale. All the Yukawa couplings
were taken as ($\sqrt{4 \pi}$) at the unification scale. 
We see from(\ref{fp}) that in such a scenario $M_Q$ becomes too   
large to be experimentally interesting. In this section we will 
go beyond this approximation and move away from the fixed point
scenario up to the degree to which correct electroweak breaking is
obtained. This is our initial reason to reduce the values of the 
couplings accordingly. Later we will see that such a reduction has 
stronger basis as it is welcome from the point of view of charge and 
color conservation.

In addition to this the complete fixed point scenario always leads to a
large value of $\tan\beta$. If we want to consider ranges of values of
$\tan\beta$ the initial values for the Yukawa couplings related to the 
bottom-tau sector ($x_f$, $x^\prime_f$) should be tuned. This causes
difficulty with our numerical subroutines. We will rather bypass this by 
taking a bottom-up approach taking $\tan\beta$ as a free input parameter 
at low energy. The initial Yukawa couplings are taken such that they 
reproduce the experimental values of  $m_b(m_b)=4.4$ GeV and 
$m_\tau=1.777$ GeV.
\begin{table}[t]
\begin{center}
\[
 \begin{array}{|c|c|c|c|}
\hline
    & \tan\beta=5 & \tan\beta=30 & \tan\beta=50\\
\hline
v_\lambda/v_S      & 0.325   &   0.317    &  0.255       \\
x_d(M_X),\,x_\tau(M_X)& 0.013,\,0.029   & 0.096,\,0.212& 0.449,\, 0.908 \\
z_q(M_X),\,z_q(M_Q)& 0.133,\,0.351   & 0.131,\,0.343& 0.086,\, 0.253 \\
z_u(M_X),\,z_u(M_Q)& 0.355,\,0.621   & 0.338,\,0.604& 0.207,\, 0.469 \\
z_d(M_X),\,z_d(M_Q)& 0.218,\,0.621   & 0.214,\,0.604& 0.155,\, 0.469 \\
k_2(M_X)           & 0.932             & 0.940         & 0.297          \\
k_1(m_Z),\,k_2(m_Z)& 0.242,\,0.158   & 0.220,\,0.172  & 0.149,\, 0.113  \\
\hline
\widehat{\alpha}_3 & 0.124          & 0.123         & 0.120          \\
M_X                &  16.94          & 16.93          &  16.88          \\
\alpha_X           &  0.252          &  0.247         &  0.226         \\
\hline
 \end{array}
\]
\end{center}
\caption{Values of Yukawa couplings and the ratio $v_\lambda/v_S$
satisfying the fermion masses. We have taken $M_Q=3 $ TeV, $m_t=175 $ GeV,
$m_b=4.4$ GeV, $m_{\tau}=1.7 $ GeV. The predictions for $\hat{\alpha}_3$, 
$M_X$ and $\alpha_X$ are also included. EWSB is guaranteed for these sets 
of Yukawa couplings. These values will be used to get the superparticle 
spectrum given in table \ref{table3}}
\label{table2}
\end{table}

Now we will describe our procedure of pinning
down  the Yukawa couplings for the
EWSB mechanism. In table
\ref{table2} we present sets of  values of $\tan \beta$ 
and the corresponding fitted values of
the Yukawa couplings. We also give the solutions for $\alpha_3$,
$M_X$ and $\alpha_X$. While fitting we have chosen the following boundary
conditions at the unification scale 
\begin{eqnarray}
x_f&=& x^\prime_f\,,\;\;\; y_f = z_f\,, \nonumber \\
z^\prime_q&=& z^\prime_l\,,\;\;\; z_u=z_\nu\,,\;\;\; z_d=z_e \,.
\end{eqnarray}
The remaining Yukawa couplings are
taken at $\sqrt{4 \pi}$. Fitted 
values of $\hat{\alpha}_3(m_Z)$
and $\alpha_X$ decrease   with $\tan \beta$. The actual
values can be found in table \ref{table2}. In the large 
$\tan\beta$ region the
Yukawa couplings related to the bottom-tau sector are
larger and we get the limit of all Yukawa couplings in the
fixed point region (section 2). Thus we have found sets of Yukawa
couplings(table \ref{table2}) giving the correct EWSB and $m_Z$
and shows that the mechanism works for ESSM as well.

Now we describe the supersymmetry spectrum of ESSM including
EWSB. The results are reported in table \ref{table3}. We have calculated
the masses of the Higgs scalars including the mixing between the 
doublets and the singlet $H_\lambda$. 
We have  calculated the masses of  squarks and sleptons,
gauginos (chargino neutralino gluino) while guaranteeing a correct
EWSB. In the first row we give the mass parameters required to obtain
the spectrum. The constraint of correct EWSB leads to an effective $\mu$
parameter in the TeV range. We note that in the case of $\mu \sim 1\, $
TeV
and low values of $M_2$ and $M_1$ the gaugino-higgsino mixing is almost
negligible. The heavier
chargino and neutralinos are almost higgsinos. The heaviest
neutralino is almost a singlino. The spectrum is not very different
from that
with an assumption of neglecting the mixing and requiring
a higgsino mass of few TeVs.
\begin{table}[t]
\begin{center}
\[
 \begin{array}{|c|c|c|c|}
\hline
m_i (GeV)  & \tan\beta=5 & \tan\beta=30 & \tan\beta=50\\
\hline
M_0,\,M_{1/2}    &  0,\, 1115   &  0,\, 1083 & 0,\,  952  \\
A_0              & -2768             &   -1922     &  -1800          \\
v_\lambda        &  4828            & 4964        &  6389         \\
\mu\equiv k_1(m_Z) v_\lambda         & 1169    & 1091  &  953 \\
B\equiv k_2(m_Z) v_\lambda+A_1 (m_Z) &  297     &  34  &    4 \\
\hline
m_{S_1}          &    64           &   73   &    81    \\
m_{S_2},\,m_{S_3}& 1342\,,1425 &  1057,\, 1490& 439,\, 1171 \\
m_{a_1},\,m_{a_2}& 1351,\, 958  &  1051,\, 1458& 436,\,  1455 \\
  m_{H^{\pm}}    &  1345           &     1057     &  441      \\
\hline
\tilde{m}_{q_i},\,i=1,2 &  1185       &    1157       &  1042 \\
\tilde{m}_{l_i},\,i=1,2 &   536       &     522       &   466  \\
\tilde{m}_{r_i},\,i=1,2 &   355       &     345       &   305   \\
\tilde{m}_{t_1},\,\tilde{m}_{t_2}& 653,\,974 & 721,\, 879 & 680,\,750 \\
\tilde{m}_{b_1},\,\tilde{m}_{b_2}& 913,\,1074 & 840,\, 984 & 617,\,816 \\
\tilde{m}_\nu    &  634             &       610     &  538           \\
\tilde{m}_{\tau_1},\,\tilde{m}_{\tau_2} & 355,\,645 & 265,\, 624 & 204,\, 568\\
\hline
\tilde{m}_{\chi^{\pm}_1},\,\tilde{m}_{\chi^{\pm}_2}  & 87,\, 1175& 89,\, 1097 & 89,\, 960 \\
\tilde{m}_{\chi^0_1},\,\tilde{m}_{\chi^0_2}  & 56,\,90  & 58,\, 90   & 58,\,92  \\
\tilde{m}_{\chi^0_3},\,\tilde{m}_{\chi^0_4}  &957,\, 957 & 1094,\, 1094& 1168,\,1174\\
\tilde{m}_{\chi^0_5} &  1530  &  1710  &  1441    \\
     M_3        &    357   &    355  &   348     \\
\hline
 \end{array}
\]
\end{center}
\caption{ Supersymmetry mass spectrum in the ESSM for our case.
$M_Q=3$ TeV and  $M_2= 90$ GeV. The values of the mass parameters at the
unification
scale and the values at the 100 GeV scale are 
included. The Higgs masses are tree-level masses. Correct EWSB is
guaranteed for this mass spectrum. Corresponding Yukawa couplings
are given in table 2. The lightest Higgs mass $m_{S_1}$ goes above
the experimental lower bound of 95 GeV when one-loop radiative
corrections are included (table \ref{table4}).}
\label{table3}
\end{table}

The same feature of small mixing is
present in the Higgs sector as well between the doublets and the singlet.
There are three CP-even states $S_1,S_2,S_3$, two CP-odd states
$a_1,a_2$ and two charged states $H^{\pm}$.
Due to the large value of $\mu$ the Higgs masses coming mainly from the
doublets, $S_1,S_2,a_1,H^\pm$ can be thought of as $h,H,a_1,H^\pm$
Higgs scalars of the standard MSSM Higgs spectrum\cite{ellis2}. We have
found that the predominantly doublet Higgs scalars satisfy the mass
relation 
\begin{equation}
m_{S1} \ll m_{S2} \sim m_{a1} \sim m_{H^{\pm}} \,.
\end{equation}

\begin{figure}[t]
\begin{center}
\epsfxsize=8cm
\epsfysize=8cm
\mbox{\hskip 0in}\epsfbox{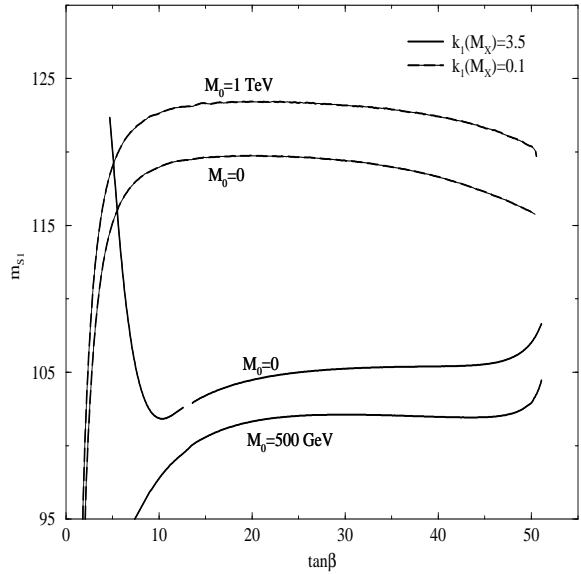}
\caption{The lightest Higgs mass including one-loop radiative corrections
from quark-squark and lepton-slepton loops. Solid (dash) lines are
computed by taking various values of $k_1$ (see legend), $M_Q=3$ TeV
and $M_2=90$ GeV.
\label{ms11}}
\end{center}
\end{figure}

Notice that the lightest (tree-level) Higgs mass quoted in table
\ref{table3} is too low to be compatible with the current experimental
lower bound of $95$ GeV. Higgs masses including the
radiative corrections from the effective Higgs potential
are given in table 4.

\subsection{Estimation of the lightest Higgs scalar mass from
effective potential}

It is well known that radiative corrections coming from the
one-loop effective potential can give an important
contribution to the Higgs masses and push the lightest
Higgs above the $m_Z$ threshold\cite{ellis2}. These corrections for
the case of the MSSM plus a singlet (NMSSM) have also been calculated
previously\cite{king}. In this
section we give a qualitative estimation of the
lightest Higgs mass. This calculation will be very
similar to similar calculations in NMSSM. The dominant corrections 
come from the top and stop
loops. We calculate a bound in the case of large $\mu$ approximation.
We use the fact that the lightest Higgs mass must be less or equal to the
minimum of one-loop improved CP-even Higgs mass matrix in
the basis $(S_1,S_2)$. In other words
\begin{equation}
m_{S1}^2 \le m_{Z}^2 ( \cos^2 2\beta + \frac{k_1^2 v^4}{m_Z^4} \sin^2
2\beta) + \frac{3}{8 \pi^2} h_t^2 m_t^2 \sin^2 \beta \left( \ln
\frac{\tilde{m}_{t1}^2 \tilde{m}_{t2}^2}{m_t^4} + \frac{(A_t+\mu \cot
\beta)^2}{(\tilde{m}_{t1}^2-\tilde{m}_{t2}^2)} \ln
\frac{\tilde{m}_{t1}^2}{\tilde{m}_{t2}^2}+ \cdots \right) \,.
\label{light}
\end{equation}
Consequently for the spectrum given in table \ref{table3} we get
\begin{equation}
m_{S1} < 120 \,{\rm GeV} \,.
\end{equation}
We have checked that upper bound is saturated when the mixing between the
doublets and the singlet is negligible that is for small values of the
coupling $k_1$.

In table \ref{table4} the numerical values of the one-loop
corrected Higgs masses including top-stops, bottom-sbottoms
and tau-staus effects are
given when $k_1(M_X)=\sqrt{4 \pi}$ and $k_1(M_X)=0.1$. Lowering the
value of the Yukawa coupling $k_1$ leaves the rest of the
spectrum (other than the Higgs scalars) practically untouched. 
The values given in table \ref{table4} are for the case
$M_0=0$. See figure 5 for the case of $M_0=1$ TeV. Very large
values of $M_0$ will imply a general increase in the
supersymmetry spectrum beyond the TeV range and may not be
experimentally interesting. This in principle can also slightly raise the
value of
$m_{S1}$ via the logarithmic dependence on stop masses in
(\ref{light}) which can be easily estimated. We skip it in
this paper. However on the brighter side this is not the case
if we take $k_1$ to be large. This can be seen  from
figure \ref{ms11} where the values with $k_1(M_X)=\sqrt{4 \pi}$ are
also plotted. In the large $k_1$ case the lightest Higgs mass
decreases with $M_0$. As we increase $M_0$ the values
of $\tan\beta < 7$ becomes incompatible with the current
experimental lower bound on $m_{S1}$.
\begin{figure}[t]
\begin{center}
\epsfxsize=8cm
\epsfysize=8cm
\mbox{\hskip 0in}\epsfbox{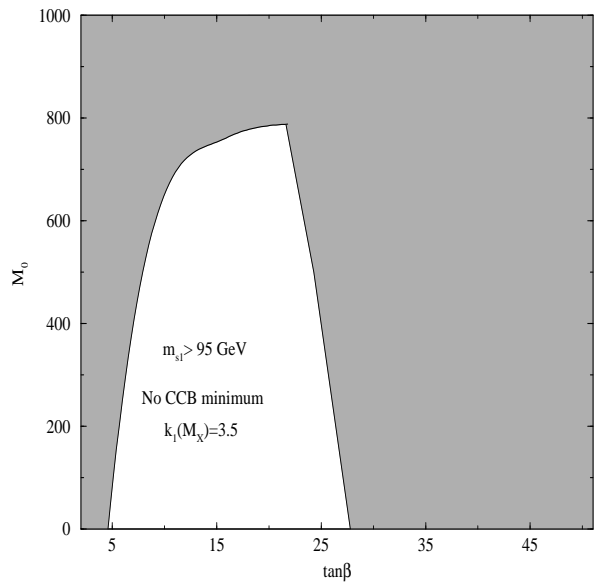}
\caption{
Taking $M_{2}=90$ GeV, we have plotted the parameter space which gives rise to 
charge and color conserving minima (unshaded region) in the $M_0 - \tan \beta $ plane.
When all the Yukawa couplings are at the fixed point
$\tan \beta$ is typically $m_t/m_b \sim 40 - 60$. Note that
there is no charge and color conserving solutions
in this case. In this figure we have taken $k^2_1 (M_X) = 4 \pi$.
\label{ccb1}}
\end{center}
\end{figure}

\begin{figure}[t]
\begin{center}
\epsfxsize=8cm
\epsfysize=8cm
\mbox{\hskip 0in}\epsfbox{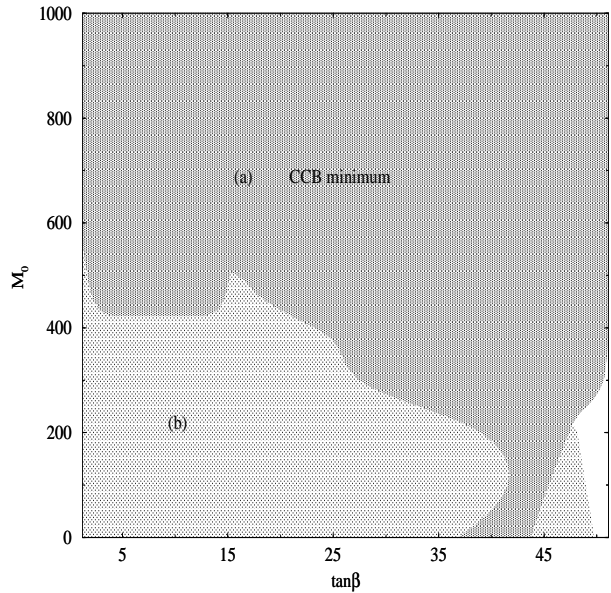}
\caption{
Taking $M_{2}=90$ GeV, we have plotted the parameter space
which gives rise to charge and color conserving minima (unshaded region) in the $M_0-\tan
\beta$ plane. Note that most of the plane is excluded. Region (b) is
due to the effect of a vector-like slepton where as (a) is due to three 
light generations only. In this figure we have taken $k_1(M_X)=0.1$.
\label{ccb2}}
\end{center}
\end{figure}

\begin{table}
\begin{center}
\[
\begin{array}{|c|c| c| c||c| c| c|}
\hline
   & \multicolumn{3}{|c||}{k_1(M_X)=3.5} & \multicolumn{3}{|c|}{k_1(M_X)=0.1} \\
\hline
m_i (GeV)  & \tan\beta=5 & \tan\beta=30 & \tan\beta=50 & \tan\beta=5 &
\tan\beta=30 & \tan\beta=50 \\
\hline
m_{S_1}          &    120    &   105    &    107  & 115 & 119 & 116
\\
m_{S_2},\,m_{S_3}& 1061\,,1423 &  1227,\, 1490& 961,\, 1171 & 1342,\, 3950
& 1176,\, 2124 & 847,\, 1439 \\
m_{a_1},\,m_{a_2}&  1090,\,934 &  1219,\, 1461& 959,\,  1455& 1342,\, 6840
& 1176,\, 3678 & 847,\, 2493 \\
  m_{H^{\pm}}    &  1061           &     1190     &  799  & 1344 & 1147 &
698     \\
\hline
 \end{array}
\]
\end{center}
\caption{Higgs masses after including one-loop radiative corrections
to the scalar potential. $M_Q=3$ TeV ($M_0=0$), $M_2= 90$ GeV
(experimental lower bound on the wino mass).}
\label{table4}
\end{table}

\subsection{ EWSB without Charge and Color breaking}

Stronger constraint on the parameter space at the unification
scale are obtained demanding that no charge and color breaking
(CCB) minima of the scalar potential is present at low
energy\cite{casas}. Note that we cannot have charge and color
breaking minima whereas we would like to have radiative EWSB and 
thereby a correct $m_Z$. Large Yukawa couplings drive
the soft supersymmetry breaking mass terms to negative values provided the
superparticles whose masses are being considered interacts via the
large Yukawa coupling under consideration. For example a large top
Yukawa coupling drives the Higgs masses to negative values
triggering radiative electroweak breaking. From(\ref{mssmyuk}) we see that
ESSM has a number of Yukawa couplings at the unification
scale which couples to the colored superpartners. Hence there is a
possibility that their large renormalization
effects may lead to masses of superparticles with weak isospin and
color quantum numbers which lead to charge and color
breaking. This will be the situation when $M_0$ and/or
$\tan\beta$ are large. Note that large $\tan \beta$ region has a large
bottom quark coupling. To get the constraint in the plane
$M_0$-$\tan\beta$ we verify  that for each set of parameters
\underbar{corresponding to a correct EWSB minimum} there is no  deeper
charge or color breaking minimum. The allowed parameter space obtained
when $k_1(M_X)=\sqrt{4 \pi}$ is given in figure \ref{ccb1}. We have
also included the constraint $m_{S1} > 95$ GeV which forbids low
values of $\tan\beta$ whereas the CCB constraint excludes large
values of $\tan\beta$. The constraint on the low values of $\tan\beta$ can
be evaded if we reduce  $k_1(M_X)$(figure \ref{ccb2}) .

Reducing $k_1$ implies allowing larger values of the couplings 
$z_f$ and $z'_c$. This turns out to be enough to affect the masses 
of the extra generation sleptons which are the eigenvalues of the 
mass matrix
\begin{equation}
{\cal M}^2 ( E'_L, E'_R )= \left( \matrix{
 \tilde{m}^2_{E_L} + z_E^2 v_\lambda^2 & z_E ( k_1 v_1 v_2 - k_2
 v_\lambda^2 + A_N v_\lambda )\cr
z_N ( k_1 v_1 v_2 - k_2 v_\lambda^2 + A_E v_\lambda & \tilde{m}^2_{E_R}
 + z_E^2 v_\lambda^2 \cr } \right)\,,
\end{equation}
where $\tilde{m}^2_{E_i}$ are the running mass parameters.
Let us discuss the slepton masses for the time being
in the context of electric charge breaking. Color breaking
can be understood very similarly. For a vector-like scale of $M_L \sim
M_Q/3 \sim 1$ TeV the
off-diagonal terms in the above mass matrix can compete with the
diagonal ones. This increases
the splitting between the eigenvalues and
even rendering one of the eigenvalues negative. Coming
back to RGE, this would be the situation
for most of the parameter space when
$k_1(M_X)=0.1$ as shown in figure \ref{ccb2}.
Squared mass parameter of a neutral slepton
becoming negative may only indicate that
it would also get a VEV, and it may affect the values
of the other VEVs. But a negative charged slepton squared
mass parameter would imply a new source of electric charge 
breaking minima in ESSM. This should be understood parallel 
to EWSB. Note that a negative squared mass parameter of the doublet 
Higgs \underbar{may not} be enough to generate a EWSB minimum. 

Thus in the small $k_1$ case  save the parameter space lost due to 
CCB a small region corresponding to large values of $\tan \beta$ is 
all that remains. We point out that there is no such constraint due 
to the vector-like spectrum for larger values of $k_1$.

\section{Conclusion}

We question the commonly accepted notion of a unified gauge
coupling $\alpha_X\sim 0.04$. If the Minimal Supersymmetric
Standard Model (MSSM) is extended by including two vector-like
families (ESSM) the couplings grow stronger than the low energy
ones due to the renormalization effects of the extra matter and
unify at a semi-perturbative scale of around $0.2$. This is
actually the only extension of MSSM containing complete families
of quarks and leptons that is permitted by measurements of the
oblique electroweak parameters on one hand and renormalization
group analysis on the other. The former restricts one to add only
vector-like families whereas the latter states that no more than
one pair of families can be added to maintain the perturbative
unitarity up to the unification scale. In ESSM the weak SU(2)
coupling grows by a factor of six at the unification scale
compared to the weak-scale value (figure \ref{ali}.a). The four
dimensional string coupling may have a similar intermediate value
which is large enough to make the dilaton stable as was
conjectured by the previous authors\cite{1}. ESSM has a unique
pattern of the Yukawa matrices which is motivated by preon
theories. The  vector-like matter and normal matter has off
diagonal Yukawa couplings whereas the normal three generations do
not have Yukawa couplings among them at all. This leads to a
see-saw like picture of the fermion masses. Below the mass scale
of the vector-like generation a hierarchical mass pattern of
chiral fermions emerge. If we fix all the Yukawa couplings to be
large at the unification scale we get  unique predictions of the
low energy fermion masses when the Yukawa couplings approach their
``quasi-infrared fixed points'' at the scale of the top quark
mass. On the contrary the renormalization effects of these
relatively large Yukawa couplings have non-trivial effects on the
unification of gauge couplings. Keeping this in mind we have also
performed the renormalization group evolution of the gauge
couplings taking into account the Yukawa effects at the two-loops.
If we assume the universality of the soft supersymmetry-breaking
parameters at the unification scale renormalization group
evolution enable us to determine the supersymmetry spectrum at low
energy quite easily. Note that due to the presence of the heavy
generations the renormalization of the superparticle mass
parameters are considerably different from that of MSSM as we
would expect. This makes ESSM distinct from MSSM from the point of
view of collider searches. The first and second generation squarks
do not have large Yukawa renormalization hence they experience
pronounced QCD renormalization which make them heavy (figure 3.b).

A further question will be to get the correct radiative electroweak
breaking. We point out that the mass of the vector-like
generations is actually linked to the electroweak symmetry breaking
mechanism by the approximate relation $M_Q \sim (z_q/k_1) \mu$. An
electroweak symmetry breaking minimum which fits the mass
of the Z boson exactly cannot be obtained in the ``quasi-infrared
fixed-point'' scenario of the Yukawa couplings if we like $M_Q$ to
be below $100$ TeVs. This is too large to be interesting experimentally.
Thus at least some of the Yukawa couplings must have smaller
values. The fixed-point scenario naturally have a large $\tan\beta$.
We show that large $\tan \beta$ region suffers from the presence
of Charge and Color breaking minima 
for any value of the vector-like scale from $m_Z$ up to $M_X$. 
So also in this case we find that to get
a global charge and color conserving minima we must give up
the assumption of all Yukawa couplings at their ``quasi-infrared fixed
point'' in the case of ESSM.

\section*{Acknowledgement}
We thank K. S. Babu, J. C. Pati and A. Rasin for discussions and 
communications. Work of BB is supported by US Department of Energy under
the grant number DB-FG02-91ER40661.

\section{Appendix}

\subsection{RGE coefficients of Yukawa couplings}

The RGE for the Yukawa couplings (at any order) are given by the
following general
expression which is perturbatively exact
\be 16 \pi^2 \frac{d\h{i}{j}{m}}{dt}=
\h{i}{k}{m}(\gamma_L)^k_j+(\gamma_R)^i_k
\h{k}{j}{m}+\h{i}{j}{m} \gamma_{H_m} \label{rge}\,.
\ee
In (\ref{rge}) \h{i}{j}{m}\ is the Yukawa coupling for the right-handed
field
$R_i$, the left-handed field $L_j$ and the scalar $H_m$. The functions
$\gamma_A$ are the anomalous dimensions for the superfield $A$\cite{11}.
We define \h{i}{j}{m}\ as the Yukawa coupling matrix when all the
superfields enter in the vertex and \hd{i}{j}{m}\ when they leave.

We have used the two-loop anomalous dimensions to evaluate the
expression (\ref{rge}) which can be splitted as follows
\be
(\gamma_A)^i_j = (\gamma_A^{1})^i_j + \frac{1}{16
\pi^2}(\gamma^{2}_A)^i_j.
\ee
The component one-loop and two-loop parts are given by the
expressions
\bea
(\gamma^{1}_A)^i_j&=&\h{i}{k}{m} \hd{k}{j}{m}-2 g_a^2 C_2(R_a)
\delta_{ij}\,,\\
(\gamma^{2}_A)^i_j&=&-\left( \h{i}{k}{m} (\gamma^{(1)}_B)^k_l
\hd{l}{j}{m}+ \h{i}{k}{m} \gamma^{(1)}_{H_m} \hd{k}{j}{m} \right)
+ 2 b_a C_2(R_a) g_a^4 \delta_{ij}
- 2 (\gamma_A^{(1)})^i_j C_2(R_a)
g_a^2 \,.
\eea
We have denoted $g_a$ as the gauge coupling $b_a$ as in the one-loop gauge
$\beta$-function
and $C_2(R_a)$ as the quadratic Casimir operator for the $R_a$ dimensional
irreducible representation. The anomalous dimensions can be
expanded in terms of four
Yukawa coupling matrices $h^m_{f}$ related to equal number of Higgs bosons
(at the unification scale they can be thought of a 10 and two singlets of
SO(10)). The index $m$ assumes the values
$m=H_1,\, H_2,\,H_S,\,H_{\lambda}$ whereas $f=u,\,d,\,l,\,\nu$. The up
sector Yukawa matrix can be re-expanded in terms of the individual
matrices which are
\bea
\mh{u}{2} &=& \left( \begin{tabular}{ccc}
  0    & $x_u$  & 0 \\
$x^\prime_q$ &   0    & 0 \\
  0    &   0    & 0
\end{tabular} \right) \,, \label{huh2}\\
\mh{u}{S}&=& \left( \begin{tabular}{ccc}
  0    &   0    & $y_u$ \\
  0    &   0    & 0 \\
 $y^\prime_q/\sqrt{2}$    &   0    & 0
\end{tabular} \right) \,, \\
\mh{u}{\lambda}&=& \left( \begin{tabular}{ccc}
  0    &   0    & 0 \\
  0    &   0    & $z_u$ \\
  0    & $z^\prime_q/\sqrt{2}$    & 0
\end{tabular} \right) \,. \label{huhl}
\eea
We will have  similar expressions for the down-quark sector and 
similarly for the leptonic sector. The normalization factor $\sqrt{2}$
avoids over counting when we sum over the index $f=u,\,d$ or $f=l,\,\nu$.

The one-loop anomalous dimensions for the quark sector are as follows

\bea
(\gam{\bar{u}}{1})_{ij}&=& 2(\mh{u}{2}\mhd{u}{2})_{ij}
+(\mh{u}{S}\mhd{u}{S})_{ij}+(\mh{u}{\lambda}\mhd{u}{\lambda})_{ij}
-\left(\frac{8}{15} g_1^2+\frac{8}{3} g_3^2\right) \delta_{ij} \,, \\
 (\gam{\bar{d}}{1})_{ij}&=& 2(\mh{d}{1}\mhd{d}{1})_{ij}
+(\mh{d}{S}\mhd{d}{S})_{ij}+(\mh{d}{\lambda}\mhd{d}{\lambda})_{ij}
-( \frac{2}{15} g_1^2 +\frac{8}{3} g_3^2) \delta_{ij} \,, \\
 (\gam{q}{1})_{ij}&=&
 (\mhd{u}{2}\mh{u}{2})_{ij}
+(\mhd{d}{1}\mh{d}{1})_{ij}
+(\mhd{u}{S}\mh{u}{S})_{ij}
+(\mhd{d}{S}\mh{d}{S})_{ij}
+(\mhd{u}{\lambda}\mh{u}{\lambda})_{ij} \nonumber\\
&& +(\mhd{d}{\lambda}\mh{d}{\lambda})_{ij}
 -(\frac{1}{30} g_1^2 + \frac{3}{2} g_2^2
     + \frac{8}{3} g_3^2) \delta_{ij} \,,\\
 & & \;\;\;\;\;\;\;\; i,j= 1,2\;
\nonumber \\
\gam{U}{1}&=& (\mhd{u}{\lambda}\mh{u}{\lambda})_{(3,3)}
+(\mhd{u}{S}\mh{u}{S})_{(3,3)}
-\frac{8}{15} g_1^2 -\frac{8}{3} g_3^2 \,,\\
 \gam{D}{1}&=& (\mhd{d}{\lambda}\mh{d}{\lambda})_{(3,3)}
+(\mhd{d}{S}\mh{d}{S})_{(3,3)}
-\frac{2}{15} g_1^2 -\frac{8}{3} g_3^2 \,,\\
& & \;\;\;\;\;\;\;\;
\nonumber \\
 \gam{\bar{Q}}{1}&=& (\mh{u}{\lambda}\mhd{u}{\lambda})_{(3,3)}
+(\mh{d}{\lambda}\mhd{d}{\lambda})_{(3,3)}+(\mh{u}{S}\mhd{u}{S})_{(3,3)}
+(\mh{d}{S}\mhd{d}{S})_{(3,3)} \nonumber\\
&& -\frac{1}{30} g_1^2 - \frac{3}{2} g_2^2-\frac{8}{3} g_3^2 \,,\\
& & \;\;\;\;\;\;\;\;
\nonumber
\eea
\bea
\gam{\bar{U}}{1} &=& \left(\begin{tabular}{cc}
$(\gam{\bar{u}}{1})_{ij}$ & 0 \\
         0                & $\gam{\bar{Q}}{1}$
\end{tabular} \right)\,, \\
\gam{\bar{D}}{1} &=& \left(\begin{tabular}{cc}
$(\gam{\bar{d}}{1})_{ij}$ & 0 \\
         0                & $\gam{\bar{Q}}{1}$
\end{tabular} \right)\,, \\
\gam{Q_u}{1} &=& \left(\begin{tabular}{cc}
$(\gam{q}{1})_{ij}$ & 0 \\
         0              & \gam{U}{1}
\end{tabular} \right)\,, \\
\gam{Q_d}{1} &=& \left(\begin{tabular}{cc}
$(\gam{q}{1})_{ij}$ & 0 \\
         0              & \gam{D}{1}
\end{tabular} \right)\,.
\eea
The two-loop contributions to the anomalous
dimensions for the quark sector are as follows
\bea
 (\gam{\bar{u}}{2})_{ij}
&=&- \left( 2 \mh{u}{2}(\gam{Q_u}{1}+\gam{H_2}{1})\mhd{u}{2}
+ \mh{u}{S}(\gam{Q_u}{1} + \gam{H_S}{1})\mhd{u}{S}
+ \mh{u}{\lambda}(\gam{Q_u}{1} + \gam{H_\lambda}{1}) \mhd{u}{\lambda}
\right)_{ij}
\nonumber \\
&&
-\left( \frac{8}{15} g_1^2 + \frac{8}{3} g_3^2 \right)
        (\gam{\bar{u}}{1})_{ij}
+\left(\frac{8}{15} b_1 g_1^4 + \frac{8}{3} b_3 g_3^4 \right)\delta_{ij}
\\
 (\gam{\bar{d}}{2})_{ij}&=&- \left(
2 \mh{d}{1}(\gam{Q_d}{1}+\gam{H_1}{1})\mhd{d}{1}
+ \mh{d}{S}(\gam{Q_d}{1}+\gam{H_S}{1})\mhd{d}{S}
+ \mh{d}{\lambda}(\gam{Q_d}{1}+\gam{H_\lambda}{1}) \mhd{d}{\lambda}
\right)_{ij} \nonumber \\
&&
-\left( \frac{2}{15} g_1^2 +\frac{8}{3} g_3^2\right)
     (\gam{\bar{d}}{1})_{ij}
+\left(\frac{2}{15} b_1 g_1^4 +\frac{8}{3} b_3 g_3^4\right) \delta_{ij} \,, \\
 (\gam{q}{2})_{ij} &=& (
 \mhd{u}{2}(\gam{\bar{U}}{1}+\gam{H_2}{1})\mh{u}{2}
+\mhd{d}{1}(\gam{\bar{D}}{1}+\gam{H_1}{1})\mh{d}{1}
+\mhd{u}{S}(\gam{\bar{U}}{1}+\gam{H_S}{1})\mh{u}{S}
+\mhd{d}{S}(\gam{\bar{D}}{1}+\gam{H_S}{1})\mh{d}{S} \nonumber\\ &&
+\mhd{u}{\lambda}(\gam{\bar{U}}{1}+\gam{H_\lambda}{1})\mh{u}{\lambda}
+\mhd{d}{\lambda}(\gam{\bar{D}}{1}+\gam{H_\lambda}{1})\mh{d}{\lambda})_{ij}
\nonumber \\
&&
-\left(\frac{1}{30} g_1^2 + \frac{3}{2} g_2^2
    + \frac{8}{3} g_3^2 \right) (\gam{q}{1})_{ij}
+\left(\frac{1}{30} b_1 g_1^4 + \frac{3}{2} b_2 g_2^4
        + \frac{8}{3} b_3 g_3^4\right) \delta_{ij} \,,\\
 \gam{U}{2} &=& -\left(
 \mhd{u}{\lambda}(\gam{\bar{U}}{1}+\gam{H_\lambda}{1})\mh{u}{\lambda}
+\mhd{u}{S}(\gam{\bar{U}}{1}+\gam{H_S}{1})\mh{u}{S}
\right)_{(3,3)} \nonumber \\
& &-\left(\frac{8}{15} g_1^2 +\frac{8}{3} g_3^2 \right) \gam{U}{1}
+ \frac{8}{15} b_1 g_1^4 +\frac{8}{3} b_3 g_3^4 \,,\\
 \gam{D}{2} &=& -\left(
 \mhd{d}{\lambda}(\gam{\bar{D}}{1}+\gam{H_\lambda}{1})\mh{d}{\lambda}
+\mhd{d}{S}(\gam{\bar{D}}{1}+\gam{H_S}{1})\mh{d}{S}
\right)_{(3,3)} \nonumber \\
& &-\left(\frac{2}{15} g_1^2 +\frac{8}{3} g_3^2\right) \gam{D}{1}
+\frac{2}{15}b_1 g_1^4 +\frac{8}{3} b_3 g_3^4 \,,\\
 \gam{\bar{Q}}{2}&=& -\left(
 \mh{u}{\lambda}(\gam{Q_u}{1}+\gam{H_\lambda}{1})\mhd{u}{\lambda}
+\mh{d}{\lambda}(\gam{Q_d}{1}+\gam{H_\lambda}{1})\mhd{d}{\lambda}
+\mh{u}{S}(\gam{Q_u}{1}+\gam{H_S}{1})\mhd{u}{S}
 +\mh{d}{S}(\gam{Q_d}{1}+\gam{H_S}{1})\mhd{d}{S}
\right)_{(3,3)} \nonumber\\
&& -\left(\frac{1}{30} g_1^2+ \frac{3}{2} g_2^2
        +\frac{8}{3} g_3^2 \right) \gam{\bar{Q}}{1}
+\frac{1}{30} b_1 g_1^4 + \frac{3}{2} b_2 g_2^4 + \frac{8}{3} b_3 g_3^4.
\eea
Two-loop anomalous dimensions of leptons can be obtained from
the above expressions with the replacements $u \rightarrow 
\nu$, $d \rightarrow e$ and $q \rightarrow l$.

We have set the Yukawa couplings for the first and second generation
to zero. They can be included by the replacement of the
numbers $x_f$, $x^\prime_c$, $y_f$ and $y^\prime_c$ by corresponding
three dimensional vectors. In this case $3 \times 3$
Yukawa matrices given in equations (\ref{huh2}-\ref{huhl}) will
become $5 \times 5$ matrices.

The one-loop and two-loop anomalous dimensions for the Higgs scalars are
as follows

\noindent (a) one-loop anomalous dimensions
\bea
 \gam{H_1}{1}&=& 3 Tr[\mh{d}{1}\mhd{d}{1}]
+ Tr[\mh{e}{1}\mhd{e}{1}]+k_1^2
-\frac{3}{10} g_1^2 - \frac{3}{2} g_2^2 \,,\\
 \gam{H_2}{1}&=& 3 Tr[\mh{u}{1}\mhd{u}{1}]
+ Tr[\mh{\nu}{1}\mhd{\nu}{1}]+k_1^2
-\frac{3}{10} g_1^2 - \frac{3}{2} g_2^2 \,,\\
 \gam{H_S}{1}&=& 6 Tr[\mh{d}{S}\mhd{d}{S}+\mh{u}{S}\mhd{u}{S}]
+ 2 Tr[\mh{e}{S}\mhd{e}{S}+\mh{\nu}{S}\mhd{\nu}{S}]+k_3^2 \,,\\
 \gam{H_\lambda}{1}&=&
  6 Tr[\mh{d}{\lambda}\mhd{d}{\lambda}+\mh{u}{\lambda}\mhd{u}{\lambda}]
+ 2 Tr[\mh{e}{\lambda}\mhd{e}{\lambda}+\mh{\nu}{\lambda}\mhd{\nu}{\lambda}]
+ 2 k_1^2 + k_2^2.
\eea
\noindent (a) two-loop contributions to the anomalous dimensions
\bea
 \gam{H_1}{2}&=& -3 Tr[\mh{d}{1}\gam{Q_d}{1}\mhd{d}{1}
                      +\mhd{d}{1}\gam{\bar{D}}{1}\mh{d}{1}]
                 - Tr[\mh{e}{1}\gam{L_e}{1}\mhd{e}{1}
                      +\mhd{e}{1}\gam{\bar{E}}{1}\mh{e}{1}] \nonumber\\
             &&  -k_1 (\gam{H_1}{1}+\gam{H_\lambda}{1}) k_1
-\left(\frac{3}{10} g_1^2+\frac{3}{2} g_2^2\right) \gam{H_1}{1}
+\frac{3}{10} b_1 g_1^4+\frac{3}{2} b_2 g_2^4 \,,\\
 \gam{H_2}{2}&=& -3 Tr[\mh{u}{1}\gam{Q_u}{1}\mhd{u}{1}
                       +\mhd{u}{1}\gam{\bar{U}}{1}\mh{u}{1}]
                 -  Tr[\mh{\nu}{1}\gam{L_\nu}{1}\mhd{\nu}{1}+
                       \mhd{\nu}{1}\gam{\bar{\nu}}{1}\mh{\nu}{1}] \nonumber\\
             &&  - k_1 (\gam{H_2}{1}+\gam{H_\lambda}{1}) k_1
-\left(\frac{3}{10} g_1^2+\frac{3}{2} g_2^2\right) \gam{H_2}{1}
+\frac{3}{10} b_1 g_1^4+\frac{3}{2} b_2 g_2^4 \,,\\
 \gam{H_S}{2}&=& - 6 Tr[\mh{d}{S}\gam{Q_d}{1}\mhd{d}{S}
                        +\mhd{d}{S}\gam{\bar{D}}{1}\mh{d}{S}
                        +\mh{u}{S}\gam{Q_u}{1}\mhd{u}{S}
                        +\mhd{u}{S}\gam{\bar{U}}{1}\mh{u}{S}]\nonumber\\
& &              -2 Tr[\mh{e}{S}\gam{L_e}{1}\mhd{e}{S}
                       +\mhd{e}{S}\gam{\bar{E}}{1}\mh{e}{S}
                       +\mh{\nu}{S}\gam{L_\nu}{1}\mhd{\nu}{S}
                       +\mhd{\nu}{S}\gam{\bar{\nu}}{1}\mh{\nu}{S}]
 \nonumber  \\
& & -k_3\gam{H_S}{1} k_3,\nonumber\\
 \gam{H_\lambda}{2}&=&
  -6 Tr[\mh{d}{\lambda}\gam{Q_d}{1}\mhd{d}{\lambda}
     +\mhd{d}{\lambda}\gam{\bar{D}}{1}\mh{d}{\lambda}
+\mh{u}{\lambda}\gam{Q_u}{1}\mhd{u}{\lambda}
     +\mhd{u}{\lambda}\gam{\bar{U}}{1}\mh{u}{\lambda}] \nonumber\\
& & -2 Tr[\mh{e}{\lambda}\gam{L_e}{1}\mhd{e}{\lambda}
     +\mhd{e}{\lambda}\gam{\bar{E}}{1}\mh{e}{\lambda}
+\mh{\nu}{\lambda}\gam{L_\nu}{1}\mhd{\nu}{\lambda}
     +\mhd{\nu}{\lambda}\gam{\bar{\nu}}{1}\mh{\nu}{\lambda}]
\nonumber  \\
& & - k_3\gam{H_\lambda}{1} k_3.
\eea
}
\subsection{Two-loop formulae of threshold corrections in step-function 
approximation.}
The two-loop coefficient for the RGE of the gauge coupling
$b_{ij}$ and $a_i^{(k)}$ including the threshold corrections
are listed here. $\theta_i$ is the step function. The
mass thresholds are denoted by
the index $i=\tilde{w}, \tilde{g},
\tilde{q}_L,\,\tilde{u}_R,\,\tilde{d}_R,\,
\tilde{l}_L,\,\tilde{e}_R,\,\tilde{h},\,H$ and similarly
the vector-like thresholds $i=Q,\,L$. Here $n_f$ is the number of chiral
generations and
$n_V$ is the number of extra generations.

\noindent (a) Gauge contribution

\bea
b_{11} &=& n_f \left( \frac{19}{15} + \left(2-\tw \right)
\left( \frac{9}{100}\tl+ \frac{18}{25}
\te+\frac{2}{75}\td+\frac{1}{300}\tq+\frac{32}{75}\tu \right)
\right) \nonumber \\
& &+ \frac{9}{50} \left( 1 + \th +\tsh \left( 1- \tw \right) \right)
\nonumber \\
& &+ n_V \left(3-\tw \right) \left( \frac{81}{100}\tll + \frac{137}{300}
\tqq \right) \,,\\
b_{12} &=& n_f \left( \frac{3}{5} + \left(2-\tw \right)
\left( \frac{9}{20}\tl+\frac{3}{20}\tq \right)
\right) \nonumber \\
& &+ \frac{9}{10} \left( 1 + \th +\tsh \left( 1- \tw \right) \right)
\nonumber \\
& &+ n_V \left(3-\tw \right) \left( \frac{9}{20}\tll + \frac{3}{20}
\tqq \right) \,,\\
b_{13} &=& n_f \left( \frac{44}{15} + \left(2-\tg \right)
\left(\frac{4}{15}\tq + \frac{6}{15} \td + \frac{32}{15} \tu \right)
\right)\nonumber \\
& &+ n_V \left(3-\tg \right) \frac{44}{15}\tqq \,,\\
b_{21} &=&
 n_f \left( \frac{1}{5} + \left(2 - \tw \right)
\left( \frac{3}{20}\tl+\frac{1}{20}\tq \right)
\right) \nonumber \\
& &+ \frac{3}{10} \left( 1+ \th + \tsh \left( 1- \tw \right) \right)
\nonumber \\
& &+ n_V \left(3- \tw \right) \left(\frac{3}{20}\tll + \frac{1}{20}
\tqq \right) \,,\\
b_{22} &=& -\frac{136}{3} + \frac{64}{3} \tw +
 n_f \left( \frac{49}{3} + \left(26 - 33 \tw \right)
\left( \frac{1}{12}\tl+\frac{1}{4}\tq \right)
\right) \nonumber \\
& &+ \left( \frac{13}{6} + \frac{13}{6} \th + \tsh \left( \frac{49}{6}-
 \frac{11}{2} \tw \right) \right)
\nonumber \\
& &+ n_V \left(25- 11 \tw \right) \left(\frac{1}{4}\tll + \frac{3}{4}
\tqq \right) \,,\\
b_{23} &=&
 n_f \left(4 +  4 \left(2 - \tg \right)\tq
\right) \nonumber \\
& &+ n_V 4 \left(3- \tg \right) \tqq \,,\\
b_{31} &=&
 n_f \left( \frac{11}{30} + \left(2 - \tg \right)
\left(\frac{1}{30}\tq +\frac{1}{15}\td+\frac{4}{15}\tu \right)
\right) \nonumber \\
& &+ n_V \frac{11}{30} \left(3- \tg \right) \tqq \,,\\
b_{32} &=&
 n_f \left( \frac{3}{2} + \frac{3}{2} \left(2 - \tg \right)\tq
\right) \nonumber \\
& &+ n_V \frac{3}{2} \left(3- \tg \right) \tqq \,,\\
b_{33} &=& -102 + 48 \tg +
 n_f \left(\frac{76}{3}  + \left(\frac{11}{3} - \frac{13}{3} \tg \right)
\left( 2\tq+ \td+\tu  \right) \right)\nonumber \\
& &+ n_V \left(40- \frac{52}{3} \tg \right) \tqq.
\eea
\noindent (b) Now the coefficients $a_i^{k}$: Let us become
careful here. We define \gamg{j}{1}\ to be the anomalous
dimensions without the gauge contributions which are as follows
\bea
\sum_k a_1^{k} &=& \frac{6}{5} \left(
     \frac{4}{3} \left(Tr \gamg{\bar{u}}{1}+\gamg{U}{1} \right)
   + \frac{1}{3} \left(Tr \gamg{\bar{d}}{1}+\gamg{D}{1} \right)
   + \frac{1}{6}\left( Tr \gamg{q}{1} +\gamg{\bar{Q}}{1} \right)
   + Tr \gamg{e}{1}+ \gamg{E}{1}
   + \frac{1}{2}\left( Tr \gamg{l}{1}+\gamg{L}{1} \right) \right.
  \nonumber \\
&& \left.+\frac{1}{2}\left( \gamg{H_1}{1} + \gamg{H_2}{1}\right) \right) \,,\\
\sum_k a_2^{k} &=& 3 \left( Tr \gamg{q}{1} + \gamg{\bar{Q}}{1} \right)
   + Tr \gamg{l}{1}+\gamg{\bar{L}}{1}
   + \gamg{H_1}{1} + \gamg{H_2}{1} \,,\\
\sum_k a_3^{k} &=&  Tr \gamg{\bar{u}}{1}+ Tr \gamg{\bar{d}}{1}
   + \gamg{U}{1}+\gamg{D}{1} + 2 \left(Tr \gamg{q}{1}+\gamg{\bar{Q}}{1}
   \right) \,.
\eea The vector-like superfields are massive at the scales ($M_Q$,
$M_L$). After the  rotation of the Yukawa matrices we get
 \bea
\mh{u}{2}(\mu < M_Q) &=& \left( \begin{tabular}{ccc} $ h_t(\mu) $&
0  & 0 \\
  0    &   0    & 0 \\
  0    &   0    & 0
\end{tabular} \right) \,,\\
\mh{u}{S}( \mu < M_Q) &=& \mh{u}{\lambda}(\mu<M_Q)=0\,, \\
\mh{d}{1}(\mu < M_Q) &=& \left( \begin{tabular}{ccc}
$ h_b(\mu) $&   0  & 0 \\
  0    &   0    & 0 \\
  0    &   0    & 0
\end{tabular} \right) \,,\\
\mh{d}{S} (\mu < M_Q)&=& \mh{d}{\lambda}(\mu < M_Q)=0\,, \\
\mh{e}{1}(\mu < M_L) &=& \left( \begin{tabular}{ccc}
$ h_\tau(\mu) $&   0  & 0 \\
  0    &   0    & 0 \\
  0    &   0    & 0
\end{tabular} \right) \,,\\
\mh{e}{S}( \mu < M_L) &=& \mh{e}{\lambda}( \mu < M_L)=0\,,
\eea
with our boundary conditions(\ref{bounc}).

%%%%%%%%%%%%%%%%%%%%%%%%%%%%%%%%%%%%%%%%%%%%%%%%%%%%%%%%%%%%%%%%%%%%%%%%

\end{document}